Monash University

# Drivers Affecting Cloud ERP Deployment Decisions: An Australian Study

This thesis is presented in partial fulfilment of the requirements for the degree of Bachelor of Information Technology (Honours) at Monash University

*By: Xinyu Zhang (27210626)*

*Supervisor: Dr Mahbubur Rahim, Dr Susan Foster*

*Year: 2019*

# Declaration

I, Xinyu Zhang, declare that this thesis is my own work and has not been submitted in any form for another degree or diploma at any university or other institutes of tertiary education. Information derived from the work of others has been acknowledged.

Signed by: 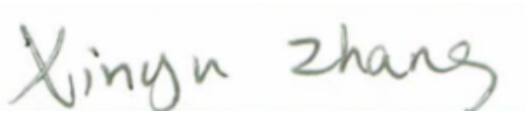

Name: Xinyu Zhang
Date: 30/05/2019

# Acknowledgements

I would like to express my sincere appreciation to my supervisor Dr. Mahbubur Rahim and my co-supervisor Dr. Susan Foster. Thank you both for being so patient with me and taking me through the whole process, thank you for guiding me and sharing your insightful opinions, and thank you for making me a confident person. You have shown me plenty of "first-times" in my academic studies: reading countless literature for the first time, understanding many professional terms for the first time, interviewing for the first time, etc. Each time, you have guided me with warm advice and patience; you have made my first research experience precious. I must say it is a great honour to be your student.

I would also like to thank the interviewee who has participated in this research. I am deeply thankful for his cooperation. Thank you for taking your precious time to conduct the interview, thank you for trusting me and being so kind and supportive. Also, I am particularly grateful for the assistance given by Dr. Vincent Lee, who provided me the chance of conducting this interview. The research would not have been completed without help from them.

Finally, I wish to thank my family. Thank you for being with me and supporting me. A special thank you to my mom who supports my studies both financially and mentally. And I would also like to thank my grandparents. I would have never accomplished this study without their understanding and encouragements.

# Abstract

*Cloud-based Enterprise Resourcing Planning (cloud ERP) is hosting an ERP system through the cloud environment. Cloud ERP is responsible for organizational business processes such as purchasing, financial, and human resource by providing a real-time infrastructure for the enterprise. With the development of technology, cloud ERP is noticed by more and more enterprises. Currently, limited researches have been conducted for cloud ERP systems in the Australian context. Furthermore, no studies have indicated how different perspectives (client company & consultant company) bring insights into the deployment decisions on cloud ERP. Hence, this research intends to understand "Drivers affecting cloud ERP deployment decisions from both client company and consultant company perspective in an Australian context". This paper identifies 31 relevant literature on cloud ERP; 79 critical drivers affecting cloud ERP deployment decisions were identified from the selected literature, and those drivers are then categorized using the Technology-Organisation-Environment (TOE) framework to develop the initial theoretical model. By conducting a Case Study Approach using a semi-structured interview and secondary resources analysis, findings are then compared to the theoretical model. As a result, an empirically validated model on drivers affecting cloud ERP deployment decisions from both client company and consultant company perspectives has been developed; this model contains 15 drivers and 7 of them are new. The theoretical and practical contributions of the findings are then outlined.*

# Table of Contents

# List of Tables

# List of Figures

# Chapter 1: Introduction

## 1.1 Introduction

In the early 1990s, the term Enterprise Resource Planning (ERP) was firstly introduced by the Gartner Group (Wylie, 1990). ERP is utilized to integrate all facets of an enterprise under a unified system (Barton, 2001). Surendro (2016) identified ERP as providing a real-time infrastructure for organizational backend systems including "purchasing, marketing, sales and inventory, procurement, financial and human resources" (p. 1). There are countless tangible benefits for enterprises to deploy ERP systems such as significant improvements in productivity, service quality, and even better business decision-making (Ngai, Law, & Wat, 2008), and intangible advantages include business integration, standardization, and improved business performance (Gargeya & Brady, 2005). Among those benefits, the most interesting one for business to implement ERP systems is to improve organizational competitive position in the market (Al-Mashari, Al-Mudimigh, & Zairi, 2003; Ehie & Madsen, 2005; Kamhawi, 2008; Ngai et al., 2008; Ram, Corkindale, & Wu, 2013; etc.); as Kamhawi (2008) once indicated "ERP is no longer viewed as merely an enabler but a tool to attain a competitive edge in the present era of globalization" (p. 318).

Despite all these benefits identified above, the implementation of ERP systems is complex and likely to fail. For instance, more than 80% of companies have failed their implementation of ERP systems in Indonesia; only 10% of companies gained success in implementing ERP systems in China (Weng & Hung, 2014). One of the significant reasons for ERP implementation failure is because of cost overruns (Ehie & Madsen, 2005; Kumar, Maheshwari, & Kumar, 2002; Scott & Vessey, 2000). Subba (2000) believed that one of the criteria for small and medium-sized enterprises (SMEs) to select ERP systems is to check the affordability.

The emergence of cloud computing has gradually eased ERP implementation failures. In 1997, the promotion and adoption of cloud computing were slow; until 2007, it increased considerably (Mei, Chan, & Tse, 2008). Many scholars (e.g. Buyya, Pandey, & Vecchiola, 2009; Mei et al., 2008; Smith, 2009; etc.) have indicated that cloud computing has the ability to offer seamless mechanisms by scaling their host services to fit into multiple data centres; yet, cloud computing is well-known for its cost-saving characteristic (Lin & Chen, 2012). One of the most significant benefits for enterprises to deploy cloud computing is financial savings; this is especially pertinent for SMEs as cloud computing provides opportunities for SMEs to acquire IT capabilities that were unaffordable while on-premise (Grossman, 2009).

Furthermore, cloud computing has three delivery models: Software-as-a-Service (SaaS), Platform-as-a-Service (PaaS), and Infrastructure-as-a-Service (IaaS) (Al-Ruithe, Benkhelifa, & Hameed, 2018). Cloud-based Enterprise Resource Planning (cloud ERP) systems are provided through the SaaS model so that system can be accessed via the cloud without any pre-installation costs by the user (Elmonem, Nasr, & Geith, 2016). Surendro (2016) proposed that "commonly used model for cloud ERP systems is the three-tier architecture" (p. 1039); this architecture has three layers and they are application layer, processing layer, and storage layer. On the client's side only lies the application layer, while processing layer and storage layer are located on the vendor's side (Surendro, 2016, p. 1039). Cloud ERP system allows users to transfer and share information in real-time due to its three-layer architecture (Surendro, 2016). Additionally, cloud ERP can be used without any requirements of IT infrastructure; any updates will be notified and handled immediately by vendors (Elmonem, 2016). Building on this discussion, the next section demonstrates the comparison between traditional ERP and cloud ERP.

## Traditional ERP vs Cloud ERP

Generally hosted on-premise ERP systems are referred to as traditional ERP. Traditional ERP systems are internally hosted within the user enterprise's infrastructure which suggests the user company is responsible for the system's updates, maintenance, as well as backup requirements (Peng & Gala, 2014). On the contrary, cloud ERP systems can be accessed through web browsers without pre-installing any applications on the user side (Elmonem et al., 2016). Table 1 below presents the comparison between traditional ERP systems and cloud ERP systems.

**Table 1**: Comparison between Traditional ERP and Cloud ERP

| Traditional ERP | Cloud ERP |
|---|---|
| 1. Accessibility (Weng & Hung, 2014)<br>2. Have high-level control over the system and more secure (Weng & Hung, 2014)<br>3. Customizable (Lenart, 2011)<br>4. Integration as "automatic data updating among related business components" (Parthasarathy, 2013, p. 178)<br>5. Own the hardware and software and manage updates (Parthasarathy, 2013)<br>6. Ongoing maintenance and support cost (Weng &Hung, 2014) | 1. Cost-efficient (Arinze & Anandarajan, 2010; Elragal & El Kommos, 2012; Mahara, 2013; Johansson, Alajbegovic, Alexopoulos, & Desalermos, 2014; etc.)<br>2. Scalable (Lenart, 2011; Nguyen, Nguyen, & Misra, 2014; Elmonem et al., 2016; etc.)<br>3. System integration with other applications (Elmonem et al., 2016; Johansson et al., 2014)<br>4. Increase business agility (Parthasarathy, 2013)<br>5. Accessibility; Availability; Fast deployment (Johansson et al., 2014) |

With regard to SMEs their move to the cloud has been exceptionally fast due to a number of factors: cost-saving, ease of access and with the advent of cloud the business is able to focus on their core functionality (Weng & Hung, 2014). Furthermore, SMEs can effectively increase business agility since the vendor manages and updates the system remotely based on a monthly fee paid directly to the vendor for this service (Lenart, 2014). In addition, SMEs usually have fewer and simpler activities "can fast deploy and utilize a constantly maintained and updated by the vendor so that this also guarantees its optimal use, ensuring their business continuity" (Johansson et al., 2014, p. 8).

As cloud ERP systems have attracted more and more attention in the market due to the various benefits identified, it is necessary to help large-scale organizations understand cloud ERP deployment drivers to engage in better decision-making. Currently, studies have been conducted on the implementation, migration journey, lifecycle, and deployment decisions of cloud ERP systems. Much of this research was conducted in developing countries such as India and Saudi Arabia, and little attention has been paid to developed countries such as the United Kingdom and the United States. In fact, no similar research has yet been reported in any scholarly literature from an Australian context. This paper seeks to bridge this gap and identifies the key drivers which are likely to affect cloud ERP deployment decisions in Australia. These factors are identified as: scalability, maintainability, accessibility, business agility, integration, etc.

In addition, cloud ERP adoption decision demands involvement and commitment of multiple stakeholders since it represents a complex technology (Surendro, 2016). In general, different stakeholders bring different perspectives in terms of their expectations on cloud ERP deployment decisions. Hence, the way various key drivers are perceived by key stakeholders for cloud ERP deployment needs attention. Yet, this aspect has remained largely ignored in the existing literature. Previously, only one paper has identified drivers on cloud ERP adoption decisions from customer and vendor points of view in an American context (Rodrigues, et al., 2016); no scholarly studies have focused on bringing different perspectives from consultant company and client company in Australian context. Hence, below presents the two research questions:

**RQ1:** What key drivers influence the deployment decision of cloud ERP systems for the Australian context?

**RQ2:** Do the viewpoints of the consultant company and the client company concerning the influence of key drivers on cloud ERP deployment decision differ?

## 1.2 Research Significance

By answering the research questions, this paper makes contributions in the following four ways.

Firstly, in a theoretical way, the drivers affecting cloud ERP deployment decisions will be categorized using the Technology-Organization-Environment (TOE) framework. In this case, it will illustrate which category is in dominance on cloud ERP deployment decisions for an Australian context. Moreover, it will also offer rich insights to explain the dominance of that category. Thereby the dominant category can then be compared with those reported in the literature for the Indian and the UK context, which would help to understand on how national contexts may have an impact on the drivers affecting cloud ERP deployment decisions.

Secondly, identification of associations among the identified drivers affecting cloud ERP deployment decisions for the Australian context will be demonstrated. This kind of associations will indicate the existence of moderating and mediating drivers. Such an analysis which has not yet been reported in any previous scholarly literature will indeed represent a contribution of knowledge by highlighting valuable insights into better understanding cloud ERP deployment phenomenon.

Thirdly, the perceptions of client company and consultant company perspectives on drivers affecting cloud ERP deployment decisions will contribute in what ways they perceive cloud ERP deployment phenomenon. Divergence or convergence in their perceptions will help the formulation of the appropriate organizational environment that must exist as a prerequisite from their viewpoints for deploying cloud ERP.

Last but not least, practically speaking, the identified drivers which affect cloud ERP deployment decisions can then be applied and adopted in different organizational context to support senior-level managers to decide whether to adopt cloud ERP systems.

## 1.3 Research Approach

This research is exploratory in nature using qualitative data analysis. It establishes a Case Study approach. Using a semi-structured interview to identify drivers on cloud ERP deployment from a client company perspective; while due to access and time limitations the secondary sources were used to analyse the consultant company perspective using client case studies found on their websites. An initial theoretical model was developed from the literature, this model will be revised once the data analysis has been completed.

## 1.4 Deliverables

The expected deliverable of this research is an empirically validated model, developed based on the Technology-Organization-Environment (TOE) framework, which presents the key drivers affecting cloud ERP deployment decisions from both client company and consultant company perspectives in an Australian context.

## 1.5 Organization of the Thesis

Overall, this thesis consists of five chapters. Chapter 1 identifies professional terms: ERP and cloud ERP, shows a comparison between traditional ERP and cloud ERP, presents the research questions, and also explains the research motivation and expected contributions.

Chapter 2 presents a systematic literature analysis. Generally speaking, 31 cloud ERP relevant scholarly literature has been identified, and the research gap has been drawn; in particular, an initial theoretical model on “Drivers affecting cloud ERP deployment decisions” has been developed.

Chapter 3 indicates the methodology which has been adopted for this research. The research model is presented in this chapter, as well as the data collection approach.

Chapter 4 describes the data analysis and provides the findings of this research. In addition, research discussion and the revised model are both demonstrated in this chapter.

Chapter 5 draws the conclusion which answers the research questions and highlights the contributions; also, limitations of this research and future research recommendations are also included in this chapter.

# Chapter 2: Literature Review

Currently, cloud ERP is still under-researched, while countless research projects have been conducted on ERP systems; few studies have focused on cloud ERP. This chapter covers two sections. Section 2.1 conducts a literature analysis of the various Cloud ERP research, establishing an approach for literature search identification, and identifying the critical drivers for making cloud ERP deployment decisions by presenting a theoretical model which is developed using the Technology-Organisation-Environment (TOE) framework under which the drivers are arranged. Section 2.2 concludes the whole chapter.

## 2.1 Literature Analysis

During the literature searching stage, the primary task is to identify cloud ERP relevant literature as much as possible. A variety of academic Information Systems databases have been utilized such as ACM digital library, IEEE Xplore digital library, Gartner, AIS electronic library, Emerald Group Publishing, ProQuest, Springer, Computer Database, and Taylor & Francis Online; other channels like Monash Library Database and Google Scholar have also been used. A set of terms have been adopted when searching for literature: "ERP, Cloud ERP, Cloud-based ERP, ERP in the cloud, traditional ERP, on-premise ERP, ERP versus cloud ERP, etc". It should be noted that no timeframe has been set during the searching process since "the concept of cloud ERP is relatively new" (Weng & Hung, 2014). Furthermore, the same reason leads to the selection of the cloud ERP relevant literature; not only A* and A level journal paper has been focused, but also other scholarly literature has been taken into consideration for the implementation of this research.

As a result, 31 papers on cloud ERP have been found. Referring to the Australian Business Deans Council (2018), the journals have all been ranked into different levels (e.g. A*, A, B, C, and other). The following table (Table 2) presents the specific number of journal papers on each level.

**Table 2:** Number of the literature identified in each journal level (**T = 31**)

| **Level** | **Journal** | **Number** | **Total** |
|---|---|---|---|
| A* - A | Australasian Journal of Information Systems | 1 | 4 |
| | International Journal of Logistics Management | 1 | |
| | Journal of Computer Information Systems | 1 | |
| | Journal of Global Information Management | 1 | |
| B | International Journal of Quality & Reliability Management | 1 | 2 |
| | Journal of Organizational Change Management | 1 | |
| C | Information Resources Management Journal | 1 | 4 |
| | International Journal of Enterprise Information Systems | 2 | |
| | Journal of International Technology and Information Management | 1 | |
| Conference | Cloud Computing Technologies, Applications and Management | 1 | 8 |
| | EuroSymposium on Systems Analysis and Design | 1 | |
| | Information Systems and Computer Networks | 1 | |
| | IT Convergence and Security | 1 | |
| | Pacific Asia Conference on Information Systems | 2 | |
| | Pre-ECIS 2014 Workshop" IT Operations Management" | 1 | |
| | Proceedings of 2013 Summer Computer Simulation Conference | 1 | |
| Other | Business Administration Dissertations | 1 | 13 |
| | Cloud Computing | 1 | |
| | Continued Rise of the Cloud | 1 | |
| | FAIMA Business & Management Journal | 1 | |
| | Future Computing and Informatics Journal | 1 | |
| | Future Data and Security Engineering | 1 | |
| | Information Development | 1 | |
| | Information Technology in Environmental Engineering | 1 | |
| | International Journal of Communication Systems | 1 | |
| | International Journal of Electrical and Computer Engineering | 1 | |
| | International Journal of Innovation, Management and Technology | 1 | |
| | Journal of Enterprise Resource Planning Studies | 1 | |
| | Quality | 1 | |
| **Total** | | **31** | |

With investigation, only 4 out of 31 pieces of literature were ranked as A* and A, 2 papers were ranked as B, and 4 papers were ranked as C. Besides, 8 papers were identified from conference proceedings; 13 papers were from other approaches such as book chapters (e.g.

Chandrakumar & Parthasarathy, 2014; Parthasarathy, 2013), and other unranked journals. The table (Table 2) also indicates that besides the International Journal of Enterprise Information Systems has two cloud ERP related papers as well as the Pacific Asia Conference on Information Systems, and other journals only have one relevant paper. Moreover, cloud ERP related literature is mainly presented in "Other" levels rather than A* - A, B, or C journal levels.

To better understand cloud ERP phenomena, the investigation on "Numbers of relative literature from 2009 to 2019" has been conducted, and Figure 1 below presents the outcome.

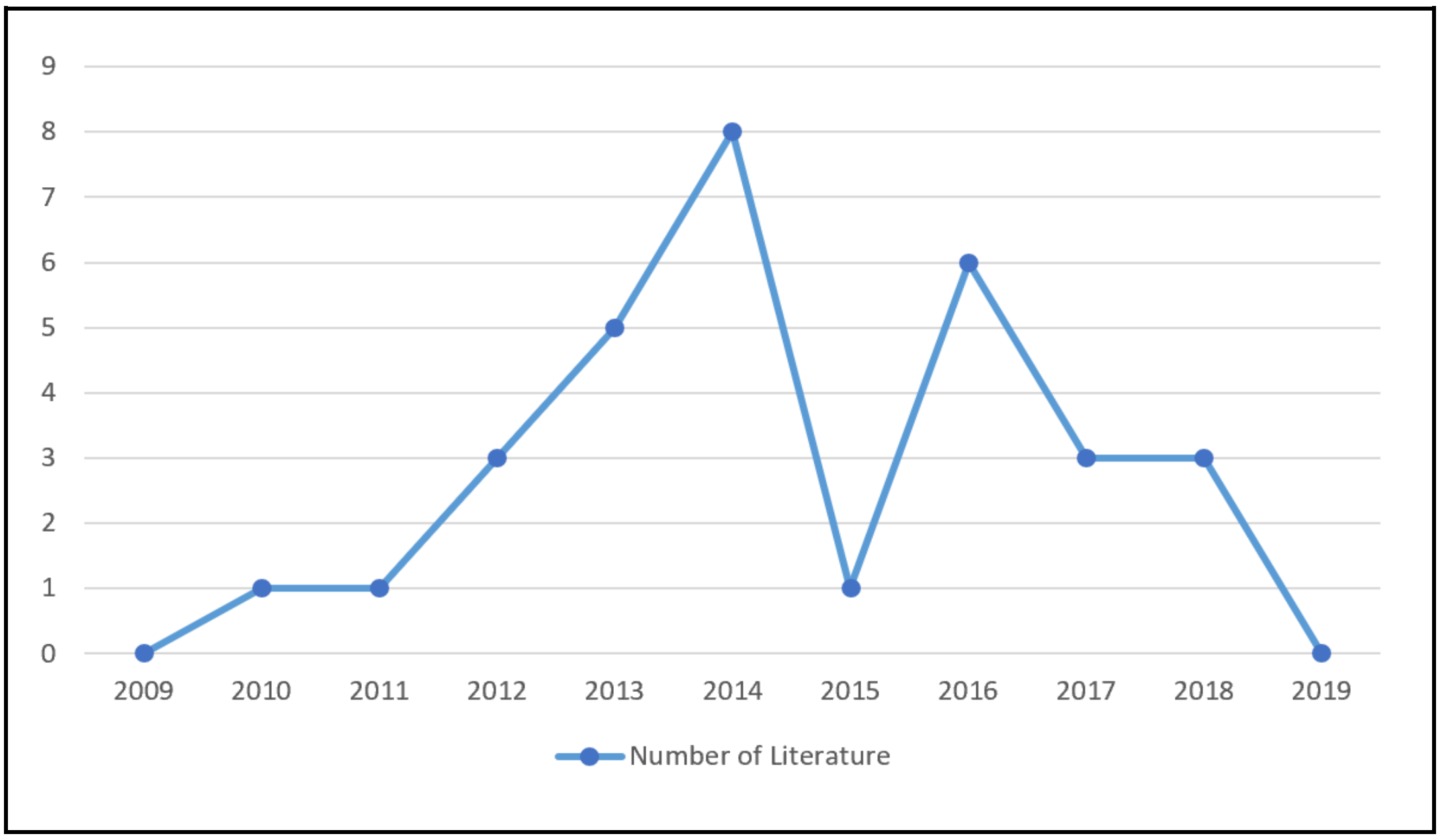

**Figure 1:** Number of Cloud ERP Relative Studies from 2009 to 2019 (**N** = **31**)

This chart (Figure 1) shows that cloud ERP relevant studies have been mostly active in 2014, which has eight related papers; then the number of literature papers has decreased dramatically in 2015, which only has one corresponding paper. However, there was no scholarly literature on cloud ERP before 2009 as well as in 2019.

After identifying the cloud ERP relative papers and knowing the peak year is 2014, another step has been conducted to categorize the identified 31 literature. The next section discusses the specific analysis of the literature. Firstly, the investigation of the research focuses on cloud ERP will be presented under section 2.1.1 Major Themes. Secondly, section 2.1.2 states the drivers on cloud ERP deployment decisions. Thirdly, section 2.1.3 presents drivers selection process. Lastly, section 2.1.4 includes the initial theoretical model on "Drivers affecting cloud ERP deployment decision" which is developed based on the selected drivers identified in section 2.1.3.

### 2.1.1 Major Themes

Overall, these 31 literature articles can be grouped into 5 main categories regarding different research focus: 1. benefits and drawbacks of cloud ERP systems, 2. migration journey from traditional ERP to the cloud, 3. lifecycle, 4. implementation or deployment critical factors of cloud ERP, and 5. other cloud ERP relative studies. Table 3 below indicates the number of literature papers for each of the category.

**Table 3:** Number of papers on cloud ERP in different journal levels by categories (**T = 31**)

| **Category** | **Journal Levels** | | | | | **Total** |
|---|---|---|---|---|---|---|
| | A* - A | B | C | Conference | Other | |
| Benefits / Drawbacks | 1 | - | - | 1 | 4 | **6** |
| Migration Journey | 2 | - | 1 | - | - | **3** |
| Lifecycle | - | - | - | 1 | - | **1** |
| Implementation / Deployment | - | 2 | 3 | 4 | 6 | **15** |
| Other | 1 | - | - | 2 | 3 | **6** |
| **Total** | **4** | **2** | **4** | **8** | **13** | **31** |

Based on Table 3, most of the literature (15 out of 31 papers) was conducted on the topic of "critical factors on cloud ERP implementation or deployment decisions", and only one paper discusses "cloud ERP lifecycle". Details on the categorization for each of the 31 literature are shown in Appendix A.

In general, 21 out of the 31 literature papers were selected to implement this research. In order to identify key drivers affecting cloud ERP deployment decisions; only two categories from the above list have been focused, and they are: 1. benefits and drawbacks of cloud ERP systems, and 2. implementation and deployment critical factors of cloud ERP. The detailed information (including the literature title, author, and publication date) of the 21 selected literature can be viewed in Appendix B.

Each of the 21 papers has been investigated by filling in a coding protocol (shown in Appendix C). The coding protocol includes three sections: section A lists the paper information including author's name, paper title, and publication year; section B shows the research characteristics such as data collection methods, theories used, size of organizations, the country where the study was conducted, and from which perspective the study was done, etc; section C depicts the deployment drivers.

By depicting the coding protocol for 21 papers, the research gap was drawn and presented in the below section.

## Research Gap

This section presents the research gap identified from literature protocol analysis, and the research gap can be illustrated in two aspects. Firstly, from the coded protocol, it has suggested that no relative scholarly focus has ever been applied to identify drivers affecting cloud ERP deployment decisions in the Australian context; hence the investigation on the distribution of cloud ERP relative studies by country has been conducted. Figure 2 below provides the distribution of cloud ERP relative literature in different countries.

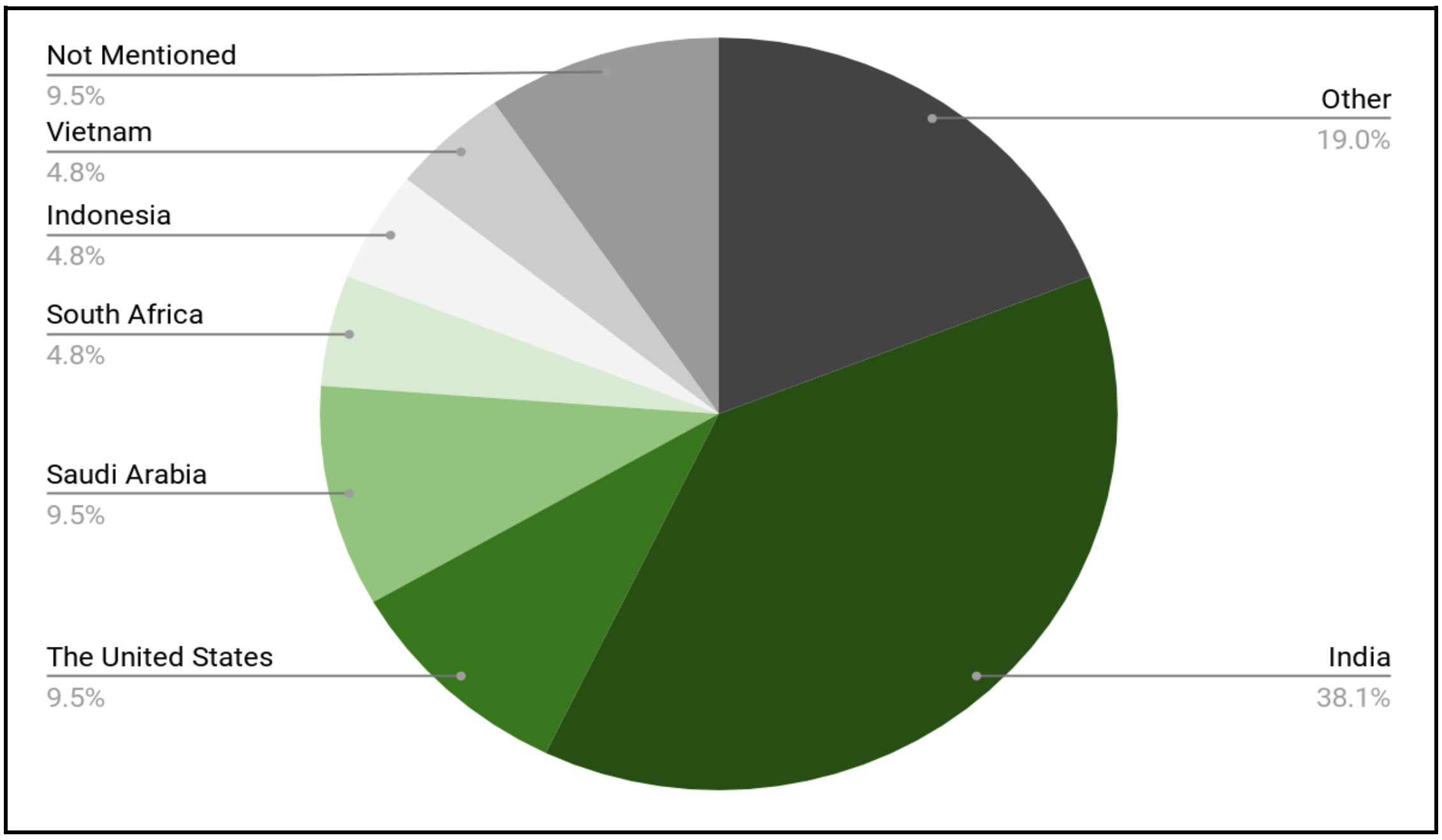

**Figure 2:** Distribution of Cloud ERP Deployment Studies by Country

This pie chart (Figure 2) illustrates that 38.1% of scholarly research (e.g. Appandairajan, Khan, & Madiajagan, 2012; Gupta & Misra, 2016; Mahara, 2013; Parthasarathy, 2013; Peng & Gala, 2014; etc.) was conducted in India which means that 8 out of 31 papers were done in an Indian context; Vietnam (Nguyen et al., 2014), Indonesia (Surendro, 2016), and South Africa (Scholtz & Atukwase, 2016) all have 1 relative paper in each of the contexts and the percentage is 4.8%; while little attention has been paid in the Saudi Arabian context (AlBar & Hoque, 2017) and in the American context (Mezghani, 2014).

Secondly, the 21 coded protocol also illustrates another research gap is that there was only one study conducted from vendor and customer perspectives (Rodrigues et al., 2016); yet, no studies have focused on the consultant company and client company perspectives in terms of drivers affecting cloud ERP deployment decision-making.

To conclude, this research explicitly determines to understand drivers affecting cloud ERP deployment decisions from an organizational (client company) and client company perspective in an Australian context. After identifying the research gap, the next section presents the findings on drivers of cloud ERP deployment decisions which were identified in the 21 selected research papers.

### 2.1.2 Drivers affecting Cloud ERP Deployment Decisions

In this section, all the identified drivers have been categorized into the Technology-Organization-Environment (TOE) framework. AlBar and Hoque (2017, p. 3) suggested that the Diffusion of Innovations (DOI) theory was also suitable for studying system deployment such as cloud ERP; the DOI theory seeks to assess how, why, and at what rate new technologies are adopted (Rogers, 2010). AlBar and Hoque (2017) have developed a model to explain cloud ERP deployment factors in the Saudi Arabia context by combing the TOE framework with the DOI theory. Within their model, six TOE factors (e.g. ICT skills, top management support, and regulatory environment, etc.) and five innovation characteristics (e.g. complexity, compatibility, and observability, etc.) have been identified (AlBar & Hoque, 2017, p. 4). Another model based on the TOE framework has been developed by Low, Chen, and Wu (2011). Within their model, there were eight critical factors such as complexity and compatibility for technological context, top management support, and firm size for organizational context, and competitive pressure for environmental context, etc. However, it should be noted that the 14 factors from both models were not empirically supported by the authors.

In this case, the TOE framework was chosen as it is an organization-level theory that explains three different elements of an enterprise's context which influence systems deployment decisions (Baker, 2012, p. 232). As well, the three categories of the TOE model demonstrate the exhaustive view within an enterprise (Gangwar, Date, & Ramaswamy, 2015). The three elements are referred to as: technology, organization, and environment. According to Baker (2012), the technological context includes both technologies currently in use and technologies available in the marketplace related to the enterprise; the organizational context refers to the resources of a firm; the environmental context consists of a "structure of an industry, presence or absence of technology providers, and the regulatory environment" (Baker, 2012, p. 235). The TOE framework was adopted for this research as it is very efficient for explaining the adoption of innovation; In this case, the TOE framework is suitable for understanding the intentions of companies regards to cloud ERP deployment.

Overall, 79 drivers were identified from 21 research articles and they have been categorized using the TOE framework. Within those 79 drivers, there are 43 technological drivers such as security, integration, functionality, etc.; 30 organizational drivers such as business complexity, cost, business agility, etc.; 6 environmental drivers such as vendor reliability, vendor lock in, regulatory environment, etc. Details on the categorization as well as the provenance of each driver can be viewed in Appendix D.

The next section aims to filter the 79 drivers from a three-step approach (Rahim, 2004) in order to develop the initial theoretical model.

### 2.1.3 Drivers Selection Process

To develop a comprehensive and empirically supported model, a three-step approach (Rahim, 2004) has been adopted and applied in this scenario. Step 1 is to eliminate duplicates in nature,

Step 2 is to extract empirically supported drivers, and Step 3 step is to identify strongly supported drivers. Below section discusses these steps in detail.

STEP 1 - Duplicated driver elimination

This initial step has been taken to eliminate duplicates from the list of 79 drivers identified in Appendix C; from the list of drivers, many of them are substantially the same yet naming differently. For instance, from the technological context, “data backup (Surendro et al., 2016), easy upgrades (Lenart, 2011), upgrade & enhancement (Peng & Gala, 2014)” are all talking about the cloud ERP system’s maintainability (Parthasarathy, 2013; Chandrakumar & Parthasarathy, 2014). According to Mahara (2013, p. 87), availability can be referred to as accessibility as the ability of the system to be accessed by users; hence, they can be grouped under “accessibility”. From the organizational context, authors like (Elmonem et al., 2016; Nguyen et al., 2014; Johansson et al., 2014; etc.) have all suggested that one of the drivers is the cost reduction for enterprise to deploy cloud ERP systems; those authors have specifically indicated which cost (e.g. operational cost, running cost, upfront cost, software cost, etc.) will be reduced, yet in this case, those specific cost drivers are all be grouped into “cost”. From the environmental context, “vendor integrity and trust in vendors” are grouped into “vendor reliability”.

STEP 2 - Extract empirically supported drivers

After taking the first step to eliminate duplicates in nature for the 79 drivers, the second step has been taken to extract empirically supported drivers. It should be noted that not all drivers identified from the above step are supported by empirical studies; hence, this step is critical in order to develop the empirically supported model. Take “cost” as an example, many authors (Elmonem et al., 2016; Chandrakumar & Parthasarathy, 2014; Appandairajan, 2012; etc.) have conducted empirical data collection methods (e.g. case study, interview, literature analysis, and surveys) to backed up their viewpoints that cost reduction is a considerable driver for enterprise to shift to the cloud. Therefore, using this approach, many other empirically supported drivers have been found like security, accessibility, integration, customization, etc.

STEP 3 - Identify strongly supported drivers

After identifying empirically supported drivers, the last step is to find which drivers are strongly supported. This step has been taken to identify drivers which have been empirically supported in more than one studies which represented as strongly supported.

As a result, by following this three-step approach, an initial theoretical model, based on the TOE framework, which consists of 10 empirically supported drivers affecting cloud ERP deployment decisions has been developed and is presented in the next section.

### 2.1.4 Initial Theoretical Model:

Hence, the initial theoretical model: “Drivers affecting cloud ERP deployment decision model” based on the TOE framework is presented below in Figure 3.

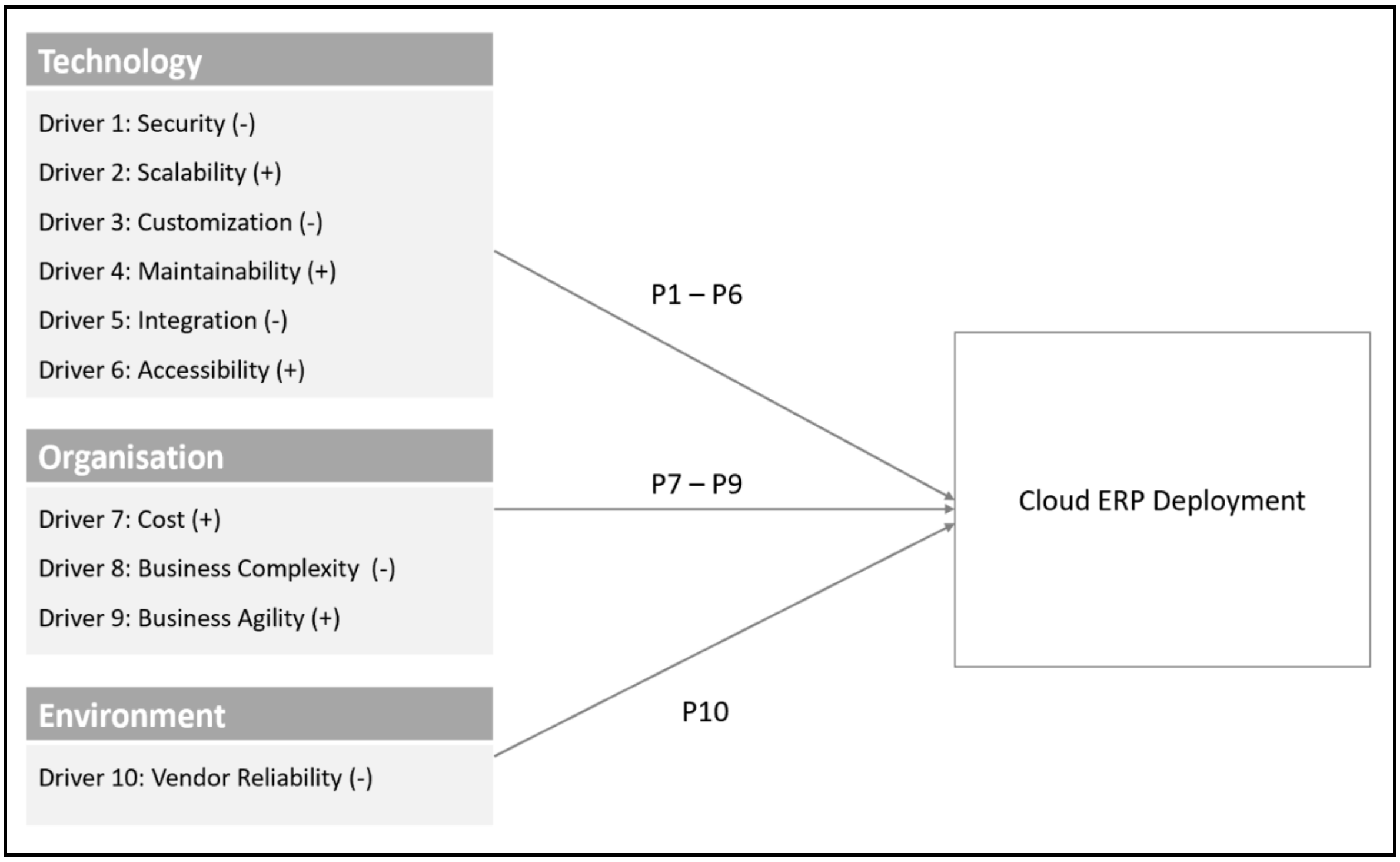

**Figure 3:** Initial Theoretical Model – Drivers affecting cloud ERP deployment decision

Besides those 10 drivers on cloud ERP deployment, this theoretical model also presents the relationship between these drivers and cloud ERP deployment decisions; whether the driver is negatively related (representing using the label "-") to cloud ERP deployment or positively related (representing using the label "+").

According to a rigorous systematic literature review, this initial theoretical model includes five negative drivers (e.g. security, customization, integration, business complexity, and vendor reliability) and five positive drivers (e.g. scalability, maintainability, accessibility, cost, and business agility) relate to the cloud ERP deployment decisions. It is notable that the technological context is the dominating category from the TOE framework; unsurprisingly technical related security issues are top concerns which can negatively influence cloud ERP deployment decisions (Arinze & Anandarajan, 2010; Garverick, 2014; Weng & Hung, 2014, etc.). In the organizational context, the cost-saving and increased business agility characteristic are the major reasons for SMEs to shift from traditional ERP to cloud ERP (Mahara, 2013; Parthasarathy, 2013; Garverick, 2014; Elmonem et al., 2016; etc.). Additionally, for SMEs, managing their own IT infrastructure is costly and exhausting; therefore, leaving the system maintenance to service providers can efficiently increase organizational agility by concentrating on core business. As for the environmental context, vendor reliability is mentioned countless times and it is interconnected with security risks; when enterprises tend to deploy cloud ERP systems, vendor reliability should also be taken into consideration.

Hence, the 10 research propositions presented in the theoretical model are discussed below.

Research Propositions

**Driver 1: Security -** Security is quite often regarded as the top one concern for enterprises when they tend to deploy cloud ERP systems (Garverick, 2014). Security within the technical aspect refers to transferring data within the cloud (Lenart, 2011), encryption and authorization with accessing data, and network security (Appandairajan et al., 2012). Arinze & Anandarajan (2010) mentioned that data security is the primary barrier of deploying cloud services in Asian countries; Mahara (2013) also indicated that "public perception on lack of security" is the major threat of cloud ERP deployment, just like what Scholtz and Atukwase (2016) suggested that "non-adopt cloud ERP enterprises are the ones which lack cloud offering knowledge" (p. 76). Yet, cloud ERP systems do expose some potential security risks. When using cloud ERP systems, the user enterprise will have to share their confidential data such as financial data and customer information with the vendor (Surendro, 2016); this will highly increase the security issues like information leakage. The most popular topic around security is data protection as well as data accessing authorization. Inadequate data protection and unauthorized data accessing can lead to specific security issues. Based on the systematic review done by Elmonem et al. (2016), the high availability on the cloud also contributes to the cloud ERP security risks. Due to those concerns, enterprises will have trouble trusting the cloud environment so that they will not be willing to deploy cloud ERP systems; therefore, the following assumption is drawn:
***P1:*** *Security is negatively related to cloud ERP deployment decisions.*

**Driver 2: Scalability** – Cloud ERP is highly elastic (Elmonem et al., 2016) and it indicates modules in cloud ERP systems can be scaled up or down rapidly depends on the organizational usage (Mahara, 2013). Garverick (2014, p. 33) mentioned that "cloud ERP systems are typically virtualized allowing for dynamic resource availability", this characteristic is highly attractive to SMEs since the enterprise can only pay for the resources which they demand rather than affording an entire software. The same idea has been proposed by Johansson et al. (2014), he indicated that scalability could be regarded as both economic benefit and business strategic benefit since cloud ERP has the ability to ensure the user company to adapt rapidly to fit the market with a minimal cost yet performs efficiently. In general, scalable cloud ERP systems can assist SMEs to reduce cost and manage resources to the corresponding usage; hence, scalability positively affects cloud ERP deployment decisions and the following proposition is presented:
***P2:*** *Scalability is positively related to cloud ERP deployment decisions.*

**Driver 3: Customization -** Customization presents as the ability for cloud solutions to fit into enterprises' requirements; "it consists of virtualization on customizing the user interface as well as editing metadata" (Nguyen et al., 2014, p. 239). Large companies are more likely to have customized modules to support their in-built systems (on-premise ERP) to ensure real-time transactions (Johansson et al., 2014); while for SMEs, cloud ERP customization is always limited since it can lead to increased runtime (Surendro, 2016). In addition, cloud ERP systems all have a standardized platform which also becomes a barrier against its customizability (Garverick, 2014) since vendors will offer user company with the standard interface. According to Lenart (2011), customization is also in major consideration for cloud ERP deployment. Furthermore, customization usually comes with high expenditure so that SMEs cannot afford such cost will not end up deploying cloud ERP with pleasure. Due to the above reasons, the following proposition is indicated:
***P3:*** *Customization is negatively related to cloud ERP deployment decisions.*

**Driver 4: Maintainability** - Maintainability refers to the system's upgrades, monitoring, backups, and disaster recovery (Gupta, Misra, Singh, Kumar, & Kumar, 2017). When deploying cloud ERP systems, the vendor will be responsible for the system's installation and maintenance (Gupta et al., 2017; Parthasarathy, 2013; Chandrakumar & Parthasarathy, 2014); this will dramatically benefit the user companies by leaving them more time to focus on their business performance rather than worrying about the system's upgrades and enhancements. Additionally, because managing the internal infrastructure as well as the system's maintenance (upgrades, backups, disaster management, etc.) is exhausting and takes a lot of time and effort; yet, using cloud ERP systems, those issues will be handled immediately by vendors so that it will directly contribute to fewer burdens for enterprise on maintenance perspective (Peng & Gala, 2014). Therefore, the statement is drawn below due to above reasons:
***P4:*** *Maintainability is positively related to cloud ERP deployment decisions.*

**Driver 5: Integration** - Integration is seamlessly integrating cloud ERP system with other applications (Weng & Hung, 2014); simultaneously, data inconsistency issues will be avoided (Surendro, 2016, p. 1040). However, integration in cloud ERP systems is hard to achieve (Garverick, 2014; Gupta et al., 2017; Saeed, Juell-Skielse, & Uppström, 2012). Integrations need to take place in multiple places such as public, private, and even hybrid clouds, as well as legacy systems (Garverick, 2014); besides, cloud ERP is standardized, users do not have full control over the system so that this can lead to difficulties for integrations. Other than that, Peng and Gala (2014) suggested that a single ERP system might not satisfy an enterprise especially for large-scale organizations, and it is pretty common for large enterprises to deploy more than one ERP platforms. ERP systems are complex and have low compatibilities which lead to low integration; yet, integration issue is much more challenging to manage in the cloud environment. Another reason is that integration cost is usually high (Peng & Gala, 2014, p. 25). Hence, organizations will have doubts about shifting to the cloud environment regarding the cloud ERP integration issue so that the hypothesis is presented below:
***P5:*** *Integration is negatively related to cloud ERP deployment decisions.*

**Driver 6: Accessibility** - Accessibility refers to the ability for users to access the system at any time anywhere from any device (Surendro, 2016; Lenart, 2011; Mahara, 2013; Nguyen et al., 2014). Cloud ERP systems have the ability for organizations and individuals to access to the platform on the vendor's side through the Internet (Mahara, 2013). In other words, the highly accessible characteristic allows the ERP applications over the cloud to be accessed, and this leads to the increase of cloud ERP usability (Elmonem et al., 2016), as well as increasing efficiency for the business performance. In general, the following assumption is indicated based on the above reasons:
***P6:*** *Accessibility is positively related to cloud ERP deployment decisions.*

**Driver 7: Cost** - Cost consists of operational cost, IT infrastructure cost, training cost, and maintenance cost (Mahara, 2013). Compared to traditional ERP systems, cloud ERP systems can reduce all the costs mentioned above which leads to the reduction of Total Cost of Ownership (Garverick, 2014). Cloud ERP definitely reduces the capital expenditure since it does not require user companies to pay for the upfront expenditure on building an IT infrastructure; all they need to do is simply buy the license to access the system (Elmonem et al., 2016). Because of the low entry cost, it is quite suitable and accessible for SMEs to deploy cloud ERP systems; SMEs are usually sensitive about cost issues so that cost-efficient can be a massive advantage for them to adopt cloud ERP systems. Additionally, not only the upfront investment is reduced, but also the maintenance costs are reduced (Mahara, 2013). Another benefit around cost for cloud ERP systems is cost-efficiency will not affect the software's license model; in other words, SMEs can obtain the same system performance benefits as large-scale companies if they are choosing the same cost model (Arinze & Anandarajan, 2010). Therefore, the following proposition is shown:
***P7:*** *Cost is positively related to cloud ERP deployment decisions.*

**Driver 8: Business Complexity** - According to Gupta et al. (2017), business complexity refers to "dealing with a huge volume of data, high load on bandwidth, encryption and decryption of data, as well as a large number of data storage server space within an enterprise" (p. 1067). Cloud ERP systems, though quite willing to adjust to changes for real-time transactions, are unable to manage across multiple business lines; complex organizations probably cannot be satisfied with what cloud ERP offers in terms of functionalities and customizations (Chandrakumar & Parthasarathy, 2014). Although cloud ERP system is complex, the functions it provides are quite limited since the system has a standardized platform; this might be the reason why SMEs usually adopt cloud ERP, since SMEs usually have fewer activities to handle compare to large-scale companies (Rodrigues et al., 2016). Hence, as the business grows more complex, cloud ERP systems will have less chance to be deployed so that the statement is indicated below:
***P8:*** *Business Complexity is negatively related to cloud ERP deployment decisions.*

**Driver 9: Business Agility** - The major benefit brought by deploying cloud ERP systems is the increasing of business agility. Business agility can be regarded from two perspectives; one is from the technical level, and another perspective is taken from the business level. For the technical level, Cloud ERP offers fast download time, rapidly adjust to changes, highly accessible, as well as no maintenance of IT infrastructure, this series of characteristics have essentially gained more efficient system performance than traditional ERP. Looking those benefits brought by cloud ERP from the business perspective, all of those strengths allow user enterprises to focus more on their core business (Arinze & Anandarajan, 2010; Garverick, 2014; Parthasarathy, 2013). Therefore, due to these reasons, cloud ERP is not only usually deployed by SMEs but also by large enterprises nowadays; this leads to the presentation of the following proposition:
***P9****: Business Agility is positively related to cloud ERP deployment decisions.*

**Driver 10: Vendor Reliability** - When deploying a cloud ERP system, the selection of vendor is quite critical. In cloud ERP systems, client companies will have to store their sensitive data (e.g. financial data and personnel data) on the vendor's side which might expose some security issues. In the cloud environment, users themselves do not have full controls over the system; they access and view what vendors provide, so that vendor integrity is highly important in this case. "Vendor reliability is related to security issues and also relates to system availability" (Arinze & Anandarajan, 2010); this leads back to the data security issues. There is no 100% foolproof solution in the current digital era as long as there are people involved, so that vendor reliability can be an negative driver when thinking about deploying cloud ERP. Hence, the following proposition is indicated:
***P10:*** *Vendor Reliability is negatively related to cloud ERP deployment decisions.*

The next section presents the conclusion to this chapter.

## 2.2 Conclusion

In conclusion, this chapter identifies the literature scope and further analyses these literature studies by filling in coding protocols for 21 selected papers. Then, the research gap was drawn from two aspects. Besides, the theoretical model is developed on "drivers affecting cloud ERP deployment decisions", and this model consists of 10 drivers. Additionally, the 10 drivers' definitions and corresponding propositions are presented in this chapter as well. The next section will discuss the methodology utilized for this research.

# Chapter 3: Methodology

This research adopts an exploratory approach using qualitative data analysis as its main data gathering approach. Multiple data gathering sources are adopted which include interview with client companies and secondary sources analysis on consultant company's websites focusing on cloud ERP deployment decision-making. In general, this chapter concretely explains the research methodology in two sections. By outlining the research design model, section 3.1 has specifically listed the flow for this research, as well as demonstrating the case study approach on data collection methods for both client company perspective and consultant company perspective; section 3.2 lastly concludes the entire chapter.

## 3.1 Research Design Model

This research has been accomplished in five stages as presented in the research design model shown in Figure 4 below.

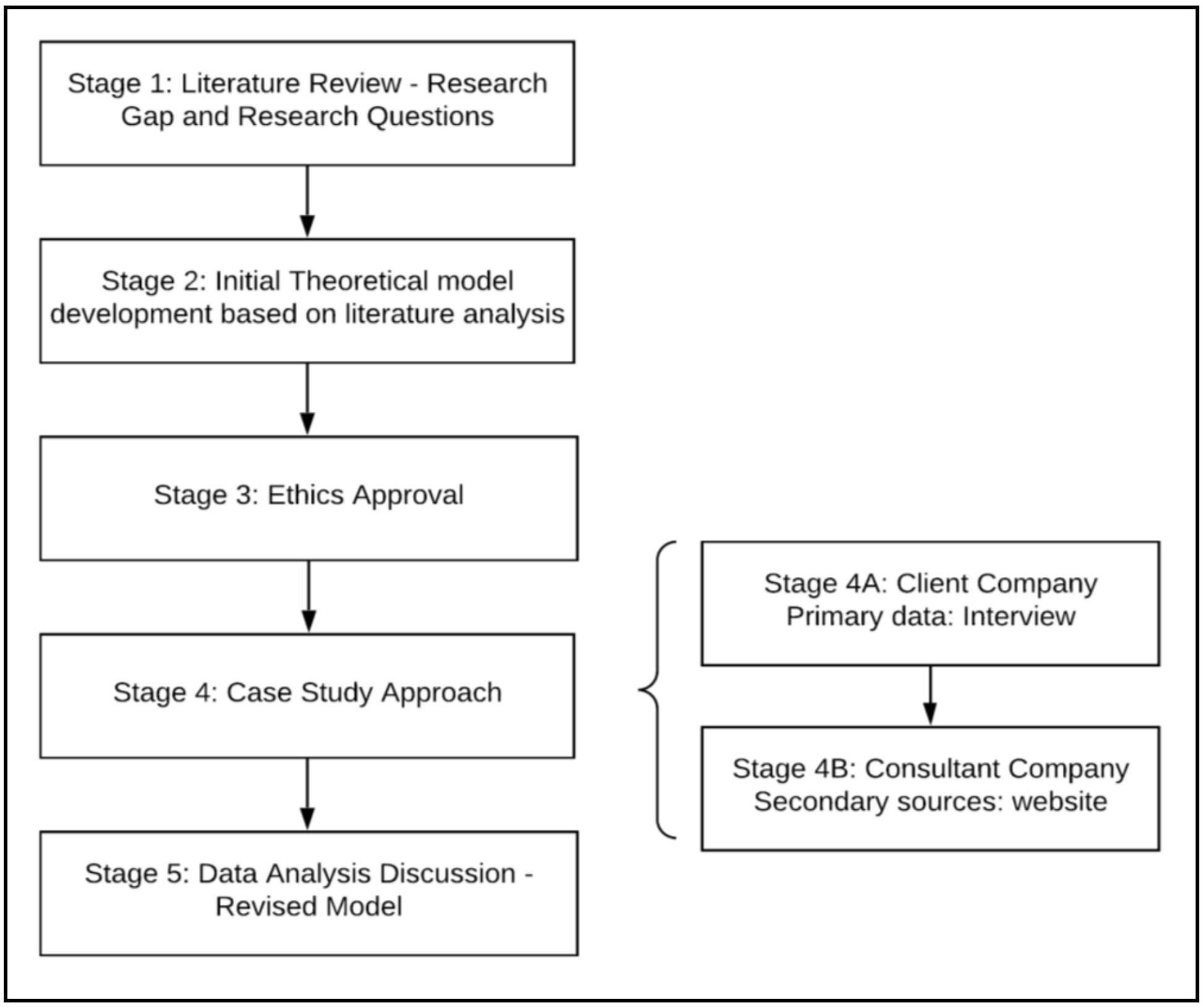

**Figure 4:** Research Design Model

This entire research has five stages. Stage 1 is to view literature identify research gaps so that two research questions have been identified. Stage 2 is the theoretical model development on cloud ERP deployment decisions based on the literature analysis. In Stage 3, the ethics approval has been applied to get permission from Monash University Human Research Ethics Committee so that interviews can be conducted. In Stage 4, a Case Study approach has been processed; an interview was done in Stage 4A, and secondary resources analysis has proceeded in Stage 4B. The entire Stage 4 aims to collect data on drivers affecting cloud ERP deployment decisions from different perspectives for better comparing with the initial theoretical model developed in Stage 2. In Stage 5, a data analysis discussion is included and the revised model which contains two perspectives has also been presented.

### 3.1.1 Stage 1 - Literature Review: Research Gap

This research starts with the literature review process. In this stage, a systematic literature review on cloud ERP deployment decisions has been conducted. A set of search terms such as cloud ERP, ERP, ERP in the cloud, cloud-based ERP have been used. Academic databases like IEEE Xplore digital library, AIS electronic library, Emerald Group Publishing, ProQuest, Springer, Taylor & Francis, Monash Library, and Google Scholar have all been used. The literature scope contains no time frame constraints; not only A* and A journal levels have been focused, but also B, C, and other scholarly literature have also been considered. As a result, 31 papers on cloud ERP have been identified.

After literature identification, the research gap was identified where no relevant studies on cloud ERP deployment have been done in an Australian context, and no studies have presented client company and consultant company perspectives for this topic. Hence, the two research questions were stated.

### 3.1.2 Stage 2 - Initial Theoretical Model Development

By categorizing the 31 literature (identified in Stage 1) on research focus, five groups of cloud ERP relative studies have been addressed: benefits and drawbacks of cloud ERP, migration journey, lifecycle, implementation and adoption critical factors, and other relevant studies. 21 papers on “cloud ERP benefits and drawbacks as well as the implementation and adoption critical factors” have been focused and utilized to develop the initial theoretical model.

The model was developed based on the Technology-Organization-Environment (TOE) framework through a three-step approach (Rahim, 2004). The three steps are: identifying all drivers for cloud ERP deployment decisions, eliminating duplicates in nature, and identifying strongly empirically supported drivers. To sum up, by using this three-step approach, the theoretical model was developed, and it consists of 10 empirically supported drivers identified by multiple studies. This model was designed to compare with the research findings gathered from interviews and secondary resources.

### 3.1.3 Stage 3 - Ethics Approval

This research follows strict and rigorous ethical guidelines and the ethics application was submitted through the Monash University Human Ethics Committee (see Appendix E) and the consent form is attached in Appendix F. With the approved ethics approval (see in Appendix G), the interview was then conducted. To further ensure the ethical perspective for this research study, the interviewee’s name will not be mentioned in this paper, and their companies are referred throughout as Case company A and B. In addition, the collected data from interviews will only be accessible to the researchers.

### 3.1.4 Stage 4 - Case Study Approach

This section presents the case study approach for both primary and secondary data collection process. In order to identify drivers on cloud ERP deployment in the Australian context, top 10 industry sectors, which have the highest sales and service income from 2016 to 2017 identified from the Australian Bureau of Statistics (2018), have been targeted in this research. As a result, this research has included two industries (Energy services and Transportation) from the list to collect primary data by conducting a semi-structured interview. Moreover, due to research access and time limitations, the other two industries (Wholesale trade and Real-estate services) are using secondary data analysis through visiting the websites of the consulting company which provided cloud ERP downloaded as at services to those industry sectors. Hence, the below section talks about the data collection approach for both client company and consultant company.

#### 3.1.4.1 Client Company

By using the Case Study approach to collect primary data from the client company perspective, a semi-structured interview has been conducted on cloud ERP deployment decision-making. Fylan (2005) suggested that semi-structured interviews are making conversations based on a list of flexible questions; using semi-structured interviews to collect qualitative data can effectively develop deep understandings of the research questions. Hence, a semi-structured interview is explicitly appropriate for better understanding of cloud ERP deployment phenomena.

After determining the data collection method, a set of criteria has been set up to select interview participants. The participants in this research have to fulfil the following criteria: 1. They must be from Australian enterprises; 2. They must be senior-level managers who have involvement in systems deployment decision-making; 3. Their organizations must have already adopted or in consideration on adopting cloud ERP systems. As a result, the interviewee was selected as a guest lecturer for Monash University. He works for case company B (public transportation company) as a Chief Architect. After communicating with him, he welcomed the opportunity to participate in this research, and an interview was conducted in his office. After conducting the interview, email communications were also included to verify the interview transcript.

In order to conduct the interview, the interviewee was provided with an explanatory statement, consent form duly signed and an interview protocol (which was designed based on the initial theoretical model) prior to the interview. The interview questions are presented below:

1. From your experience, what are the drivers that you think were considered important when making cloud ERP deployment decisions for your organizational context?
2. Are there any technological factors (e.g. data security, scalability, integration, customization, accessibility, maintainability) that you think were considered important by your organization when making cloud ERP deployment decision?

3. Are there any specific business-related factors (e.g. cost issues, business agility, business complexity) that you think were considered important by your organization when making cloud ERP adoption decision?
4. Are there any environmental factors (e.g. vendor reliability) that you think were considered important by your organization when making cloud ERP adoption decision?

The interviewee was involved in two companies: an Australian energy company (Case company A) and an Australian state-level public transportation company (Case company B); both companies are listed in the top 10 highest sales and income industry sectors from the Australian Bureau of Statistics (2018). Below section presents the organizational background of the two case companies.

#### Case A: Energy Company

The case company is an Australian public company which provides services of generating and retailing of energies such as gas and electricity for residential and commercial use. The energy company has over 180 years of history and over 3 million customers. According to the Australian Bureau of Statistics (2002), a company which has more than 200 employees is seen as large-scale companies. In this case, the company Case A is a large-scale organization with more than 3,700 employees and their net income is above $1.000 billion in 2018.

The interviewee, was initially the enterprise architecture (EA) manager, has worked in case company A for more than two years; then he joined the transportation company (case company B) as a Chief Architect. As an EA manager, he was the technical adviser and in a major drive to make recommendations on system selections. In general, the interview lasted around 60 minutes to discuss the cloud ERP deployment drivers; as a result, seven drivers were identified in case company A’s context.

The below section talks about the transportation company (case company B)’s background.

#### Case B: Transportation Company

Case company B was formed in 2012; it is dedicated to offering daily transportation services to the public. The case company B is also a large-scale organization with more than 500 employees in 2018. In addition, this company is a state-level organization which performs under government jurisdiction.

The Chief Architect was interviewed who has recently joined this public transportation company. In general, five drivers on cloud ERP deployment decisions were indicated for this organizational context which will also be presented in the next chapter.

In general, by analysing the interview data, drivers identified by the interviewee have been compared to the drivers in the initial theoretical model identified from the literature; a table which contains the identified cloud ERP deployment-related drivers will be presented from the client companies’ perspective in the next chapter. Besides, the two case companies were also included to participate in the cross-case comparison (Williamson & Johanson, p. 180) of cloud ERP deployment phenomena which will also be presented in chapter 4.

The next section discusses the data collection method used for the consultant company perspective by analysing secondary resources on drivers affecting cloud ERP adoption decision.

#### 3.1.4.2 Consultant Company: Annexa

The Case Study approach has been adopted to collect data from secondary sources focusing on drivers affecting cloud ERP deployment decisions using a consultant company perspective. This approach can effectively provide insight into the consultant companies' opinions in terms of their support to help companies to adopt cloud ERP services. Given the access limitations to consultant companies to obtain primary data (even multiple consultant companies have been contacted, no replies were received), plus the time constraints, multiple secondary resources (e.g. websites including online journals and online publications) have been used.

By assessing consultant company's websites to collect secondary data on cloud ERP deployment, Annexa, the consultant company was selected. Since this research tries to identify drivers affecting cloud ERP deployment decisions in Australia, not only client companies have to be Australian-based, but also the consultant company has to be Australian-based as well. Annexa is an Australian-based consultant company which locates in Melbourne and Sydney. Annexa provides the best technical solutions for its customers by offering Oracle NetSuite which is the top #1 cloud-based business software suite in Australia (Annexa, 2019). Within Annexa's official website, there are many case companies to whom they have offered cloud ERP adoption experiences. In order to select the appropriate case companies to implement this research, the case companies would need to meet the following requirements: 1. be Australian based organizations or have subsidiaries in Australia; 2. adopted cloud ERP systems; 3. their cloud migration experience can be found online through consultant companies' websites. As a result, Annexa was chosen as they offered two companies case studies: Jak Max and REA Group. Jak Max is a wholesale trade supplier and Rea Group is a Real Estate Agency. Both industries are included in the top 10 highest sales and income list (Australian Bureau of Statistics, 2018). These case companies are discussed in the next section.

##### Case C: Jak Max Pty Ltd

Jak Max is a wholesale trade supplier of outdoor power equipment such as chainsaw spare parts and cut-off saw spares and have commenced business in 2006 in Melbourne. Jak Max is well-known of its high quality and up-to-dated products. Moreover, Jak Max always ensures that "they have offered dealers with the latest products from the world's leading manufacturers" (Jak Max, 2015, para. 3).

Jak Max was looking to expand nationwide and has chosen Annexa consultant company to develop an advanced ERP system with high functionality, high profitability, and high productivity for their business. By analysing the secondary data, four drivers were identified relative to cloud ERP deployment decision-making for Jak Max.

##### Case D: REA Group

The real-estate agency company, REA Group, was founded in 1995 with the headquarters in Melbourne; it offers global online real-estate advertising. REA Group has more than 1400 employees working across three continents and has more than 20 brands across 6 countries. Moreover, REA Group has above $807 million in revenue in 2018 (REA Group, 2004).

According to the Annexa’s official website, REA Group was requesting a scalable ERP solution to efficiently increase their system operations in order to support global expansion in 2014. Due to the high reputation of REA Group, there is a good deal of journals, papers, and online newspapers have reported this event. Generally, there are five factors identified related to cloud ERP adoption decisions for REA Group context.

In general, by investigating secondary data to get cloud ERP deployment-related drivers and comparing them to the theoretical model, another cross-case comparison table from consultant companies’ perspective has been developed and presented in chapter 4.

To conclude, in this research, the case study approach includes the semi-structured interview to collect primary data from client companies’ perspective, and secondary data analysis has been adopted to gather data from consultant company perspective due to time and access limitations.

### 3.1.5 Stage 5 - Data Analysis and Discussion

After developing the two cross-case comparison tables on cloud ERP deployment from both consultant company perspective and client company perspective, an aggregated revised model is developed and will be demonstrated in the next chapter. This finalized model will include cloud ERP adoption drivers categorized in technological, organizational, and environmental contexts; in addition to that, the model will depict two perspectives to illustrate drivers affecting cloud ERP deployment decisions.

The next section demonstrates the conclusion of this chapter.

## 3.2 Conclusion

To conclude, this chapter has narrated two sections for explaining the methodology used for this research. The first section describes the research model including 5 stages and the last section concludes the overall chapter.

Due to the time and access constraints, case study approach is divided into two parts, where the client company perspective is done through a semi-structured interview, while the consultant company perspective can only be done through secondary data analysis approach.

The next chapter presents the data analysis outcomes and discussions.

# Chapter 4: Data Analysis

This chapter presents the findings of the semi-structured interview and secondary sources. This chapter is divided into 3 major sections: section 4.1 presents the drivers on cloud ERP deployment decisions from client companies' perspective; section 4.2 describes the drivers from the consultant company's perspective; section 4.3 provides the insightful discussion of the outcomes and a revised model will be demonstrated; lastly, section 4.4 provides a conclusion to this chapter.

## 4.1 Client Company

The 60 minutes semi-structured interview was conducted with the Enterprise Architecture (EA) Manager (Case Company A) and Chief Architect (Case Company B). The interview questions are again presented below:

1. From your experience, what are the drivers that you think were considered important when making cloud ERP deployment decisions for your organizational context?
2. Are there any technological drivers (e.g. data security, scalability, integration, customization, accessibility, maintainability) that you think were considered important by your organization when making cloud ERP deployment decision?
3. Are there any specific business-related drivers (e.g. cost issues, business agility, business complexity) that you think were considered important by your organization when making cloud ERP adoption decision?
4. Are there any environmental drivers (e.g. vendor reliability) that you think were considered important by your organization when making cloud ERP adoption decision?

Since this interview is semi-structured and the questions were asked, we allowed the interviewee to lead the interview by sharing his opinions on cloud ERP deployment decisions. Hence, the following part presents the interview outcomes on drivers affecting cloud ERP deployment from client company perspective.

### 4.1.1 Case A: Energy Company

According to the Enterprise Architecture (EA) Manager from the case company, "*the enterprise was looking forward to developing the S/4HANA system hosted by SAP in the cloud environment rather than continuing to use the SAP ERP solutions*". By answering the interview questions, there were seven drivers considered as important when case company A was making cloud ERP deployment decisions. It should be noted that there are drivers identified from the interviewed data analysis that are outside the theoretical model, and these drivers are identified as ND (New Driver).

***Technological Drivers:***

**Security** - There are always debates about security risks for cloud computing adoptions, and security is always considered as the top #1 concern for cloud ERP deployments. However, the EA manager has expressed a mixed opinion in terms of the security factor for his organizational context. The EA manager indicates that "*security can be seen from technical security risks and human-related security risks*". The case company is "*confident with their technical-related security risks because they have already had a lot of experience with cloud computing*". The company has a "*rigours cloud certification process and auditing process to deal with sensitive data so that the cloud environment is even more secure than on-premise infrastructure*". On the contrary, human-related security risks such as data leaking and third-party issues cannot be fool-proofed. Yet, this kind of "*human-related security risks does not apply uniquely to cloud ERP systems but also can happen to on-premise ERP*". Hence, security can be regarded as a mixed driver relates to cloud ERP deployment.

**Integration** - Integration, as system and process incorporation capability, is a major driver for the case company, when considering cloud ERP deployment. According to the EA manager, "*all modules like HR and Finance are currently in the same environment so that integration and latency issues are not involved, and the case company is confident with their current SAP system*". However, "*when moving particular one module to the cloud environment, integration becomes a serious issue and needs to be well considered in order to prevent any negative impacts on user experience*". Hence, integration is always a major driver which negatively related to cloud ERP deployment decision-making.

**ND: Maturity of ERP systems in the cloud environment** - The leading driver for the company to decide whether they should adopt cloud ERP systems is the maturity of the cloud ERP software. This is a new driver (not existing in the theoretical model) identified by the EA manager. Case company A is a large-scale organization where each decision-making is severely critical especially when they are dealing with adopting or implementing a new system. According to the EA manager, "*it is necessary to ensure the maturity of the potential adopting cloud ERP systems, this includes whether there is already a customer base of the system in the market*". Large-scale companies should not take any risks to deploy an immature system which might lead to implementation failures or even worse. Therefore, the maturity of ERP systems in the cloud environment can be regarded as a barrier to the cloud ERP deployment decisions.

***Organizational Drivers:***

**Cost** - As mentioned by the EA manager, "cost" is another driver for them to consider the cloud ERP deployment. For the case company A, they have two options: "*one is to upgrade current on-premise SAP ERP solutions to the latest version, second is to deploy S/4HANA system in the cloud environment*". To better understand the cost difference between these two options, "*a cost comparison has been conducted; notably, the result suggests that the two options are basically cost-equivalent*". Since "*cloud ERP solutions do not provide the enterprise with a dramatic cost reduction*", a neutral attitude has been given for the cost driver.

**Business agility** - It is indicated by the EA manager that "*the case company was aiming to put their customers first and they wish to get rid of the heavy physical data centre*". "*Their current SAP ERP system was managed by themselves, and they were tired of maintaining the physical infrastructure*". "*By deploying the S/4HANA cloud system, the vendor will take responsibility* *for maintaining the infrastructure; in the meantime, the enterprise will be able to engage in focusing on its core business as well as servicing customers*" so that business agility will be increased. In general, business agility is a critical factor for the case company to deploy cloud ERP.

**ND: Change of operating model** – Change of operating model is another new driver identified by the EA manager. The case company A's current operating model involves many responsibilities in terms of operating the system, database, and software. Yet, according to the EA manger, "*deploying a new cloud ERP system will impact everyone within the organization*". It is always hard for enterprises to adopt a new system especially when their employees are already familiar and comfortable with their existing system (Gupta, 2017). Hence, "*moving to the cloud environment will indeed cause roles and responsibilities changes*". According to the EA manager, "*with this kind of operating model change, the case company does not have a specific answer on if there will be staff restructuring or staff cutting*". Therefore, the change of the current operating model is definitely a major driver of cloud ERP deployment for the case company, since they need to consider "*whether they are ready to embrace all the changes*".

***Environmental Driver:***

**Vendor reliability -** According to the EA managers, the case company "*has a tight commercial relationship with its vendors: SAP and Microsoft*". "*The vendors and the case company have contract relations in terms of legislation*". "*Vendor reliability is ensured by having security service-level agreement (SLA), data protection policy, as well as security guidelines which all prevent data leaking issues to an extent*". Hence, vendor reliability is not a concern for the case company's context.

Compared to the initial theoretical model, there are similarities as well as differences. Same drivers are found with the interview outcomes such as security, integration, cost, business agility, and vendor reliability; yet, two new drivers are indicated: maturity of ERP systems in the cloud environment and change of operating model. Drivers from the theoretical model like scalability, customization, accessibility, and business complexity, etc. are not included in the interview. Additionally, the relationships between these drivers and cloud ERP deployment decisions are also varied from the theoretical model.

Next section will demonstrate the interview outcomes identified from case company B.

### 4.1.2 Case B: Transportation Company

According to the interviewee (Chief Architect) from the case company B, "*this company has already deployed cloud-based solutions for their Human Resource (HR) functions; currently, they are in considerations on moving Financial departments to the cloud environment as well using Oracle Financial*". Overall, the chief architect has mentioned five drivers on cloud ERP deployment decisions, including the drivers for previous HR function and the drivers for future Oracle cloud-based Financial functions.

***Technological driver:***

**Security** – According to the Chief Architect, "*Case company B is looking for an independent service provider due to cost issues and difficulties in vendor relationship management*". Yet, bad choice on the selection of service providers will expose serious security issues. Additionally, "*case company B does not have many cloud experiences so that it is even harder to choose a suitable vendor*". Therefore, security is a critical driver which needs to be considered when making decisions on cloud ERP deployment.

***Organizational Drivers:***

**Cost** – It is indicated by the Chief Architect, "*Costing was the major motivation for the case company to move HR functions to the cloud*". It has been mentioned during the interview, "*their previous HR functions were developed on-premise; however, it was very costly around these services*". Since cloud ERP systems can bring cost reduction for the enterprise, the HR functions have already been shifted to the cloud environment.

**Business agility** – According to the Chief Architect, "*the nature of case company B is to provide transportation services daily to the public; hence, the company should not be an IT development house as well as an infrastructure maintenance house*". "*Those heavy maintenance duties should be left to the vendor so that the company can focus on its business solution including providing services to passengers, engaging in better customer experience, establishing more applications and features for customers, etc.; other technical layers like operating the system, database, server, network, storage should all be handed over to service providers*". By deploying cloud ERP systems, business agility will be increased dramatically for any business.

**ND: Vendor Relationship management** – The third new driver is identified in case company B's context. According to the Chief Architect, "*case company B is currently performed under the state-level providing services so that vendor relationship management is uniquely applied in this organization*". "*The relationship with the service provider is difficult to manage since it is an agency-to-agency relationship instead of a customer-to-provider relationship*". Besides, "*there are many other agency departments sharing the same solution from the service provider; the case company is not the only customer of this on-premise solution*". Since case company B usually performs under the government's strategies, "*the Oracle Financial movement to the cloud environment is the government's initiative*". Hence, the difficulty with managing the relationship with its service provider is a critical driver for the company to "*look for an independent service provider to deploy their financial functions in the cloud environment*".

***Environmental Driver:***

**ND: Political reason** – According to the chief architect, "*Case company B performs under government legislation, obligations, and strategies; hence, political issues will affect every major decision-making*". "*Private organizations have full power in terms of decision-making; yet, public organizations like case company B needs to follow a set of guidelines and instructions, and each decision will need to go through a rigorous process*". More than that,

"*the election will also affect the organization's decision-making process; major decisions like system deployments will be made after the election*". In general, political reasons can be seen as an environmental related driver which negatively affects cloud ERP deployment decisions in case company B.

In general, the five drivers identified for case B are different from case A to a great extent due to the difference in organizational essence. Compared to the theoretical model, cost factor, security, and business agility are corresponding; yet, political reasons and vendor relationship management are uniquely applied to case B context.

The next part demonstrates the cross-case comparison on cloud ERP deployment decisions from client company perspective including all the drivers identified above.

### 4.1.3 Cross-Case Comparison - Client Company Perspective

In order to directly compare the interview outcomes with the drivers identified in the initial theoretical model, Table 4 below presents a cross-case comparison for the two client companies (case company A and case company B) on drivers affecting cloud ERP deployment decision. In general, there are 9 drivers identified from client company perspective.

**Table 4**: Drivers Comparison - Client Company Perspective (N = 9)

| **Drivers** | **Evidence (Y/N)** | | **Supporting Status** |
|---|---|---|---|
| | **Case A** | **Case B** | |
| Theoretical Model-Based Drivers | | | |
| Driver 1: Security | Y | Y | Mixed |
| Driver 2: Scalability | N | N | No Evidence |
| Driver 3: Customization | N | N | No Evidence |
| Driver 4: Maintainability | N | N | No Evidence |
| Driver 5: Integration | Y | N | Negative |
| Driver 6: Accessibility | N | N | No Evidence |
| Driver 7: Cost | Y | Y | Mixed |
| Driver 8: Business Complexity | N | N | No Evidence |
| Driver 9: Business Agility | Y | Y | Positive |
| Driver 10: Vendor Reliability | Y | N | Neutral |

| New Drivers | | | |
|---|---|---|---|
| ND 1: Maturity of ERP systems in cloud environment | Y | N | Negative |
| ND 2: Change of Operating Model | Y | N | Negative |
| ND 3: Vendor Relationship Management | N | Y | Positive |
| ND 4: Political Reason | N | Y | Positive |

This table (Table 4) includes four major outcomes: 1. Whether the 10 drivers identified in the initial theoretical model are considered important for the two case companies, as whether the two case companies show evidence (e.g. yes represents as “Y”, no represents as “N”) for each drivers in the theoretical model; 2. What is the supporting status of each driver, as what is the relationship between each driver and cloud ERP deployment (e.g. “Negative”, “Positive”, “Mixed”, “Neutral”, and “No Evidence”); 3. Whether there are new drivers identified from the interview, and what are the relationships between new drivers and cloud ERP deployment decisions; 4. Which drivers are strongly supported (two companies both considered important) and which are weakly supported (only one company considered important).

Based on Table 4, there are three drivers (e.g. security, cost, and business agility) which are strongly supported as they are considered important by both companies; yet, their relationships with the cloud ERP deployment decisions are not consistent. For instance, there is no need to worry about technology-related security risks in case company A; however, case company B indicated security is a major driver when making system adoption decisions. Moreover, cost-efficient of cloud ERP is a critical driver and it is applied in case B; yet, no significant cost reduction has been identified in case company A. Therefore, a mixed opinion has been drawn in both security and cost drivers. In addition, as for business agility, both organizations have mentioned that deploying cloud ERP will lead to an increase in business agility so that a positive supporting status has been given to this driver. In this case, security and cost supporting statuses are different from the theoretical model, while business agility’s supporting status is consistent with the theoretical model.

Drivers like Integration and Vendor Reliability are weakly supported as they were only considered important by case company A, and there is no evidence showing the existence of these two drivers in case company B. It should be noted that case company A has the same opinion with the theoretical model where Integration should be negatively related to cloud ERP deployment. Yet, a differ has been shown for Vendor Reliability where the theoretical model suggested this driver should be negatively related to cloud ERP deployment, while case company A indicated it as neutral.

Moreover, other drivers from the theoretical model such as scalability, customization, accessibility, maintainability, and business complexity are not supported by any of the client companies. There is no evidence showing that these drivers have influence on cloud ERP deployment decisions in these two case companies.

Other than that, there are emergences of four new drivers, which have not been indicated in any previous scholarly literature, are mentioned by the two case companies. Case company A has identified "maturity of ERP systems in the cloud environment" and "change of operating model" where they are both negatively related to cloud ERP deployment; the other two new drivers "vendor relationship management" and "political reasons" identified by case company B are positively related to cloud ERP deployment. This could be due to the fact that this research is so current that cloud adoption issues are changing constantly, and it is difficult for the literature to keep up. In general, nine drivers on cloud ERP deployment have been identified from client companies' perspective.

The next section will present the cloud ERP deployment drivers from consultant companies' perspective.

## 4.2 Consultant Company: Annexa

Using the case study approach, many secondary sources (e.g. websites, online journals, online articles) are utilised to be analysed. By mainly analysing the consultant company's website, the following part presents the outcomes of cloud ERP deployment drivers from the consultant company's perspective using the two client companies' deployment experience.

### 4.2.1 Case C: Jak Max Pty Ltd

Annexa, the consultant company, presents a case study on their website on helping Jak Max to choose the right ERP solution to support a global expansion for the enterprise. Based on the Jak Max case experience on cloud ERP deployment from the Annexa's official website, four drivers have been illustrated and will be presented below.

***<u>Technological Drivers:</u>***

**ND: Functionality** – According to Annexa, "*one of the major reasons for Jak Max to deploy a new system is because they were struggling under a system with limited functionalities*". "*The organization was looking forward to adopting a system with greater efficiency as well as reducing unnecessary loss due to inventory mismanagement*". In addition, "*Jak Max was also facing a cumbersome reporting system which leads to low efficiency and low productivity*". Based on these reasons, Jak Max knew it was time to adopt something new to "*replace their current system so that business profitability and productivity will be increased*". Hence, Annexa has brought NetSuite OneWorld, the cloud ERP system, to Jak Max; "*with the new system, greater business efficiency has been achieved due to the various functionalities the cloud ERP supports*". Hence, functionality is a major driver for Jak Max to adopt cloud ERP.

**Scalability** – Based on the information provided on Annexa's website, "*Jak Max believed that adopting a single unified system will lead to an efficient, flexible, and scalable business*". "*NetSuite cloud-based ERP system allows the enterprise to easily and quickly adapt to market requirements*". According to Annexa (2019), "*Jak Max was requiring transforming its* *enterprise architecture to an agile and rapidly scalable platform*" (para. 8). Based on the above reasons, scalability is an essential driver for Jax Max to deploy the cloud ERP system.

**Accessibility** - As mentioned by Annexa, "*Jak Max was seeking for a unified system which allows transactions to happen simultaneously across three subsidiaries*". Since the global expansion will be achieved by adopting a cloud system, "*employees will be able to log in the system nationally or internationally at any time*". Additionally, "*high accessibility will bring increased business productivity since employees can check the stock levels whenever necessary so that stocks will always be ensured*". With the cloud ERP system, accessibility can be well guaranteed; also, high accessibility indirectly ensures business performance of Jak Max. Hence, accessibility is another driver for Jak Max to adopt cloud ERP system.

***Organizational Driver:*** Non-identified

***Environmental Driver:***

**ND: Global expansion** – According to Annexa, "*Jak Max expected to deploy a unified system to support its global expansion so that a better blueprint can be drawn for the company*". In order to do that, they need a cloud-based ERP solution which will "*allow real-time transactions to happen across multiple subsidiaries as well as seamlessly manage those subsidiaries*". NetSuite cloud-based system helps Jak Max to meet its requirements so that a sustainable global business has also been achieved. "*With the advanced functionalities, inventory control, and faster response time, the cloud ERP system has dramatically helped Jak Max to succeed with its global expansion*". Hence, supporting global expansion is a major reason for Jak Max to deploy cloud ERP systems.

Based on the consultant company's (Annexa) website, four drivers on cloud ERP deployment decisions have been identified from Jak Max's context. Two drivers (scalability and accessibility) have been found corresponding with the theoretical model, as well as their relationships with cloud ERP deployment decisions are also indicated as the same. Two new drivers are identified from Jak Max context, and they are functionality and global expansion. Functionality is not included in the theoretical model while global expansion has not been mentioned in any previous scholarly literature. These drivers will be presented again in section 4.2.3 as the consultant company's perspective on cloud ERP deployment.

The next section discusses the drivers on cloud ERP deployment decisions for REA Group.

### 4.2.2 Case D: REA Group

Annexa also presents the REA Group case study on helping the organization to choose the suitable cloud-based ERP system to improve their business efficiency as well as supporting their global business. Since the company is famous and reputable, much relevant information can be found online for in-depth understanding their intentions on cloud ERP deployment decisions. In addition to the Annexa's official website, other resources like online newspapers and journals have also mentioned about REA Group adopting NetSuite cloud ERP system.

Overall, five cloud ERP deployment drivers have been identified for REA Group and these drivers are presented below.

***Technological Drivers:***

**ND: Issues with the current system** – According to Annexa, "*the major driver for REA Group to deploy a new system is they were struggling with their previous ERP system which was Sage ACCPAC*" (Annexa, 2014). Sage ACCPAC is known as Sage 300 (Sage, 2018), which is an ERP system and mainly adopted by SMEs; and it was firstly deployed in REA Group 15 years ago (Cowan, 2016, para. 3). According to Annexa, "*the platform provided by Sage ACCPAC was lack of flexibility so that the entire business was unable to evolve and to keep up with the market requirements*". As a result, with the help of Annexa, "*26 instances of Sage ACCPAC were replaced with NetSuite OneWorld*" (Annexa, 2014). Hence, the issues with the current system is a driver for the enterprise to deploy a new cloud-based ERP.

**ND: Functionality** – It is indicated in Annexa's official website, "*Sage ACCPAC has functional constraints which cause many of the business processes were done manually so that the business performance is inefficient*". One of the examples is the "*revenue recognition calculation function is done manually so that it exposes plenty of issues such as miscalculations and low efficiency*" (Annexa, 2014, para. 5). Since Sage ACCPAC mainly supports SMEs, it is hard for the system to handle large-scale organizational activities. Due to these reasons, REA Group knew it was time to make a change so that they were looking for a system which can "*provide the organization with greater efficiency, higher productivity, and comprehensive features*". Hence, functionality is a driver for REA Group the get rid of their current system and to deploy a cloud-based ERP.

**Scalability** – "*As the business grows, the increased volume of transactions cannot be scaled appropriately due to the legacy system issues*" (NetSuite, 2016). More than that, "*the NetSuite OneWorld supports REA Group to scale efficiently to meet market requirements and expand globally without any significant costs*" (NetSuite, 2016). Therefore, REA Group takes scalability as another important factor on making cloud ERP deployment decisions.

**Maintainability** – Maintainability is an essential driver for REA Group context in terms of cloud ERP deployment. According to Annexa, "*because the enterprise has plenty of legacy platforms that are not able to upgrade automatically so that they might expose issues including technical risks and business drawbacks*"; those legacy platforms are usually slow to run and will affect business performance, and they are quite costly to maintain (Bisbal, Lawless, Wu, & Grimson, 1999). On the other side, "*deploying the OneWorld system, REA Group no longer needs to worry about the maintenance of legacy systems; in addition, the system will always be at the latest version which ensures the business performance*". Therefore, high maintainability is another reason for the REA Group to deploy the cloud ERP system.

***Organizational Driver:*** Non-identified

***Environmental Driver:***

**ND: Global expansion** – According to Annexa, "*Sage ACCPAC was no longer suitable for REA Group since the company has grown nationwide and now is a multinational (including Asia, Australia, Europe, and North America) property advertising company*" (Annexa, 2014). With their previous system, "*many of the manual processes would take several days to perform; with NetSuite OneWorld, all the transactions are happening real-time because of global consolidation*". "*In order to support the enterprise's global growth, REA Group needs a consolidating process which provides a holistic view of the entire organization*". Hence, global expansion is another driver for the REA Group to deploy cloud ERP.

In general, five drivers have been identified by Annexa from the case of REA Group. Compare to the theoretical model, two drivers are the same and they are scalability and maintainability. Global expansion and limited functionality are corresponding with the Jak Max's context. Other than that, a new factor has been identified for REA Group's context which is "issues with the current system". These five drivers will be presented in the next section to demonstrate the cross-case comparison between theoretical model and consultant company's perspective.

### 4.2.3 Cross-Case Comparison - Consultant Companies

To better compare those drivers identified in section 4.2.1 and section 4.2.2 with the theoretical model, Table 5 below presents the secondary data outcomes. The same approach used for Table 4 has been established with this table; it includes the driver's evidence existence (whether the driver is identified by the consultant company), driver's relationship with cloud ERP deployment ("Positive", "Negative", "Mixed", "Neutral", "No Evidence"), emergence of new drivers, and whether the driver is strongly/weakly supported (support by one case company or by both companies). Hence, Table 5 is presented below:

**Table 5**: Drivers Comparison - Consultant Company Perspective (N = 6)

| **Drivers** | **Evidence (Y/N)** | | **Supporting Status** |
|---|---|---|---|
| | **Jak Max** | **REA Group** | |
| Theoretical Model-Based Drivers | | | |
| Driver 1: Security | N | N | No Evidence |
| Driver 2: Scalability | Y | Y | Positive |
| Driver 3: Customization | N | N | No Evidence |
| Driver 4: Maintainability | N | Y | Positive |
| Driver 5: Integration | N | N | No Evidence |
| Driver 6: Accessibility | Y | N | Positive |
| Driver 7: Cost | N | N | No Evidence |
| Driver 8: Business Complexity | N | N | No Evidence |
| Driver 9: Business Agility | N | N | No Evidence |
| Driver 10: Vendor Reliability | N | N | No Evidence |
| New Drivers | | | |
| ND 5: Functionality | Y | Y | Positive |
| ND 6: Global Expansion | Y | Y | Positive |
| ND 7: Issues with Current System | N | Y | Positive |

According to Table 5, 6 drivers have been identified from consultant company's perspective. Compared to the theoretical model, only one driver is strongly supported as it has been supported by both case studies which is scalability; both case studies have shown a positive supporting status on this driver which means that scalability is positively related to cloud ERP deployment decisions for both organizations. In addition to that, maintainability and accessibility is weakly supported as maintainability is supported by REA Group while accessibility is supported by Jak Max. Moreover, these three drivers' relationships towards cloud ERP deployment decisions are fully supported by the theoretical model.

However, theoretical model-based drivers like security, customization, integration, cost, business complexity, business agility, and vendor reliability are not supported by any of the case companies (case company C and case company D). In general, the consultant company's perspective did not show any evidence on the existence of those 7 drivers.

Furthermore, three new drivers have emerged as shown in Table 4: functionality, global expansion, and issues with current system. It should be noted that functionality has actually been mentioned countless times in previous scholarly literature. The reason of taking functionality out from the theoretical mode is that there is always debate around cloud ERP functionality; during the three-step approach to develop the theoretical model, functionality did not meet requirement in step 2 on "identifying empirically supported drivers". Functionality is quite controversial as some studies have shown cloud ERP have limited functionalities than on-premise ERP systems (Garverick, 2014 and Johansson et al., 2014); while in this case, functionality is fully supported by both Jak Max and REA Group cases so that functionality is strongly supported. The other two new drivers are "global expansion" and "issues with current system". These two drivers are not indicated in any scholarly literature, and they are directly related to cloud ERP adoption in a positive manner; global expansion is supported by both Jak Max context and REA Group context, so it is strongly supported.

Then the next section discusses the overall findings including the primary data and the secondary data; the next section also presents the revised model to show both perspectives (client company and consultant company).

## 4.3 Discussion

This section is divided into three parts. Section 4.3.1 presents the revised model which contains both client company and consultant company perspectives, as well as the comparison between the theoretical model and the revised model. Section 4.3.2 presents the emergence of new drivers identified in the revised model. Section 4.3.3 presents the drivers comparison between client company perspective and consultant company perspective.

The theoretical model, which was developed based on the TOE framework, was used to design interview protocol as well as making comparisons with the empirical findings. Figure 3 below presents 10 drivers derived from 21 papers on cloud ERP deployment; yet, no study indicates which perspective (e.g. client company, consultant company, or vendor) the paper has taken. The theoretical model is once again presented below to better comparing with the revised model.

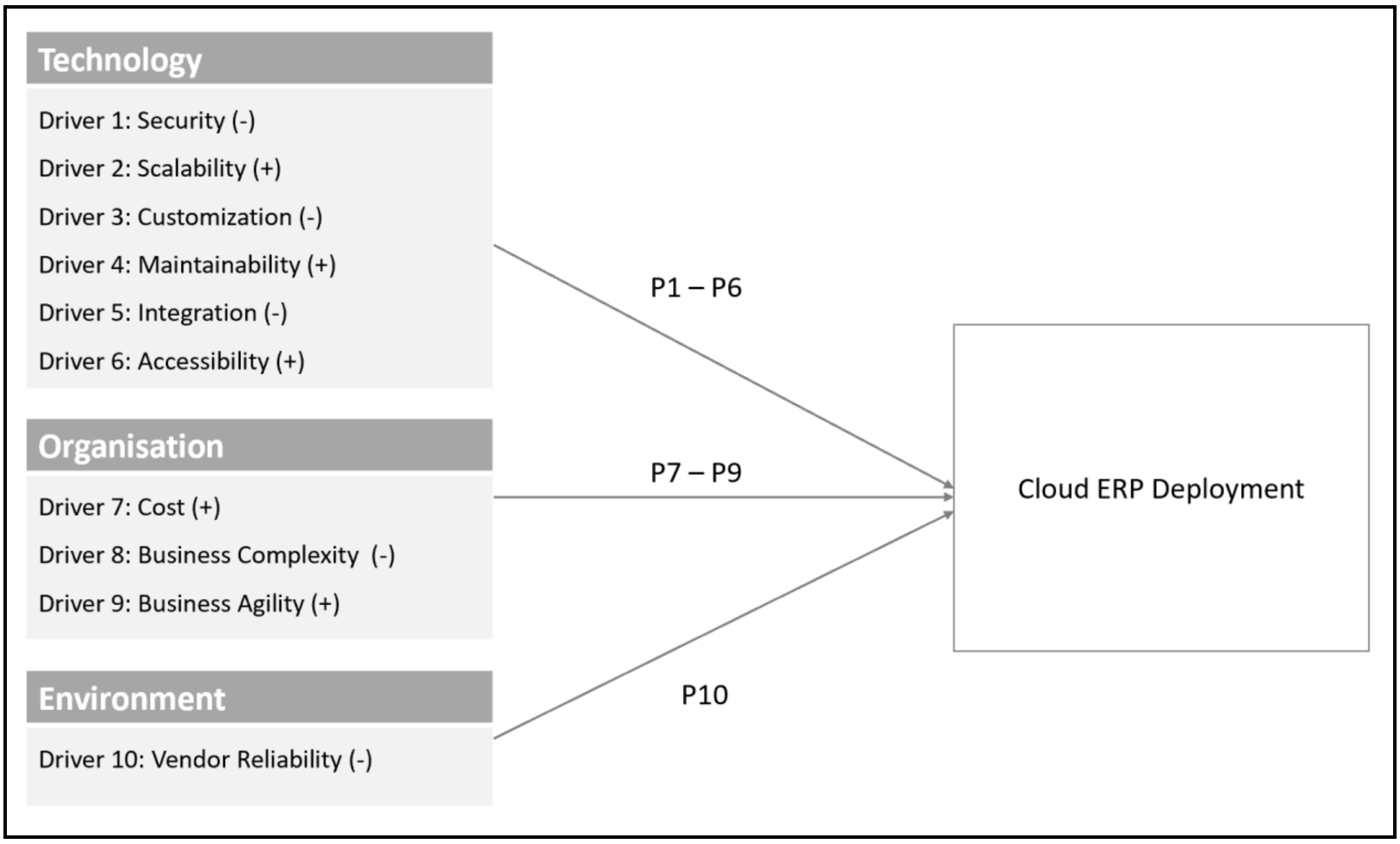

**Figure 3 Review:** Initial Theoretical Model – Drivers influencing cloud ERP deployment

Based on conducting the Case Study approach which contains the findings of the interview and the secondary resources analysis, 15 drivers on cloud ERP deployment have been identified from both client company and consultant company perspectives. The revised model, which contains those 15 drivers identified from two perspectives, is also developed using the TOE framework and is presented below in Figure 5.

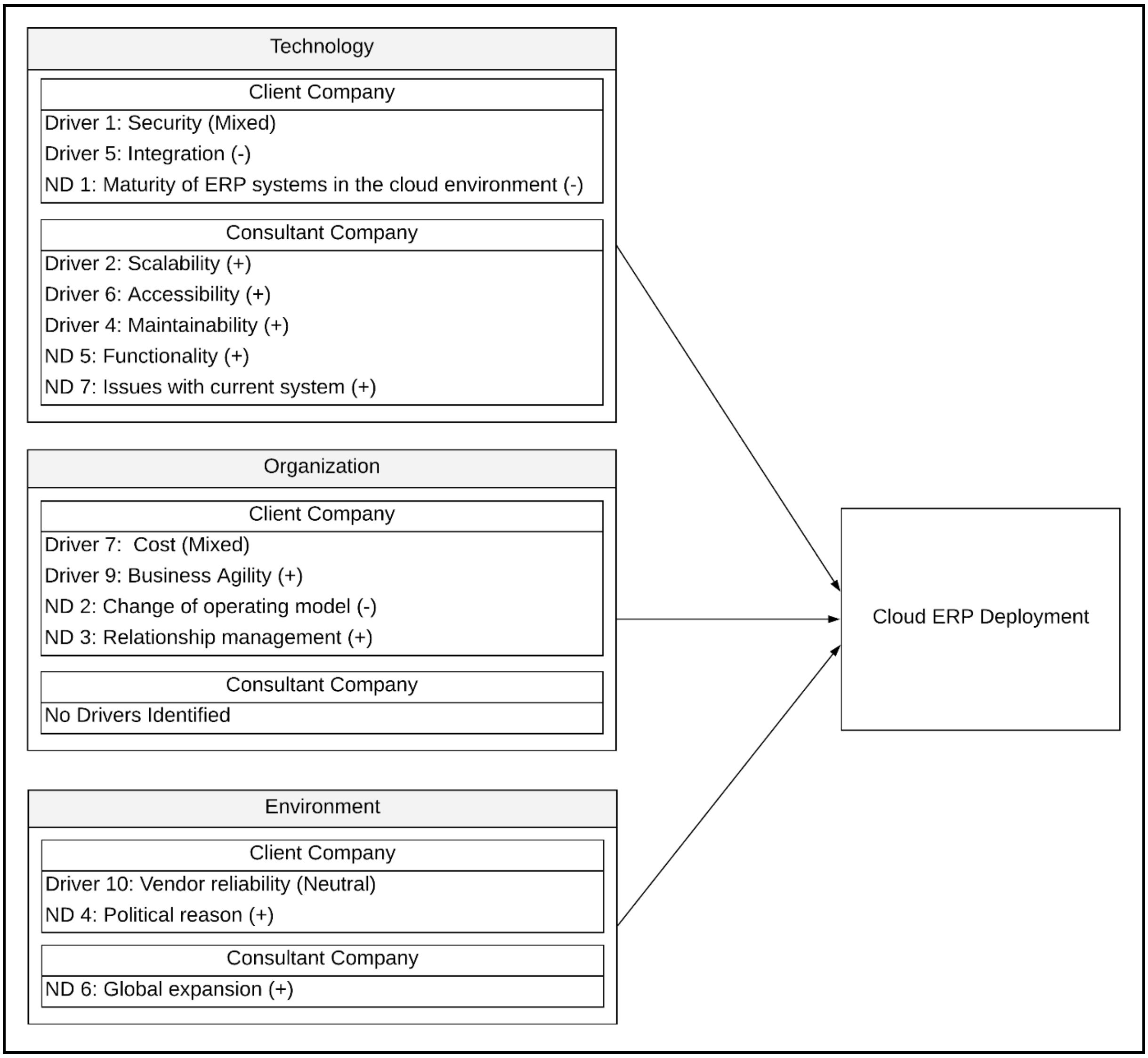

**Figure 5:** Revised Model – Cloud ERP Deployment Model with Two Perspectives (N = 15)

This revised model presents that there are 15 drivers in general; 8 drivers are from the technological context, 4 drivers are from the organizational context, and 3 drivers are from the environmental context. For each of the context, the model has taken both client company perspective as well as the consultant company perspective in terms of cloud ERP deployment decisions. This revised model contains 9 drivers from client company perspective including 5 theoretical model-based drivers and 4 new drivers; the model also includes 6 drivers from the consultant company perspective with 3 drivers from theoretical model and 3 new drivers. And, the next section discusses the comparison between the theoretical model and the revised model.

### 4.3.1 Theoretical Model vs Revised Model

By comparing the theoretical model (shown in Figure 3) with the revised model (shown in Figure 5), there are 8 out of 10 drivers from the initial theoretical model have been supported by the revised model; and these drivers are security, scalability, maintainability, integration, accessibility, cost, business agility, and vendor reliability. Other than those 8 drivers, there are 2 drivers form the theoretical model that has not been supported by the revised model, and they are customization and business complexity; this means that neither client companies nor consultant companies have shown existing evidence on these two drivers. Hence, the below section presents the proposition testing outcomes of the 8 supported drivers.

In order to test the propositions drawn in the theoretical model, Table 6 below indicates the supporting status (supporting status represents as “No evidence”, “Positive”, “Negative”, “Mixed”, and “Neutral”) of each theoretical model-based driver in the revised model.

**Table 6**: Driver Supporting Status by Two Perspectives (Theoretical Model)

| Theoretical Model-based Drivers | Supporting Status | |
|---|---|---|
| | **Client Company** | **Consultant Company** |
| Driver 1: Security | Mixed | No evidence |
| Driver 2: Scalability | No evidence | Positive |
| Driver 3: Customization | Not Supported | Not Supported |
| Driver 4: Maintainability | No evidence | Positive |
| Driver 5: Integration | Negative | No evidence |
| Driver 6: Accessibility | No evidence | Positive |
| Driver 7: Cost | Mixed | No evidence |
| Driver 8: Business Complexity | Not Supported | Not Supported |
| Driver 9: Business Agility | Positive | No evidence |
| Driver 10: Vendor Reliability | Neutral | No evidence |

According to the table above (Table 6), 5 drivers (e.g. scalability, maintainability, integration, accessibility, and business agility) are fully supported by the revised model as each of the drivers’ relationships towards cloud ERP deployment decisions are corresponding with the theoretical model. Also, both models suggest that scalability, maintainability, accessibility, and business agility are positively related to organizational decision-making on cloud ERP deployments, and integration is negatively related to cloud ERP deployments. In short, the revised model supports five corresponding propositions and they are presented below:

***P2:*** *Scalability is positively related to cloud ERP deployment decisions.*
***P4:*** *Maintainability is positively related to cloud ERP deployment decisions.*
***P5:*** *Integration is negatively related to cloud ERP deployment decisions.*
***P6:*** *Accessibility is positively related to cloud ERP deployment decisions.*
***P9****: Business Agility is positively related to cloud ERP deployment decisions.*

However, the other 5 propositions are not supported by the revised model. 3 propositions out of 5 are not fully supported since the driver's relationship to cloud ERP deployment is not corresponding with the theoretical model, and these drivers are security, cost, and vendor reliability. The below section presents the proposition outcomes of those drivers.

Security, in most cases, is a significant concern for organizations when making cloud computing adoption decisions (Lenart, 2011; Appandairajan et al., 2012; Elragal et al., 2012; etc.); while in this case, client companies have provided a mixed opinion on this factor. The case company A is confident with the cloud environment since they have plenty of cloud experiences. The interviewee mentioned that the case company A is relaxed in terms of technology-related security risks such as dealing with sensitive data; while for human-related security risks such as data leakage or third-party issues cannot be one hundred percent prevented. Hence, P1 is not supported since the revised model suggests that there is a mixed opinion on security.

The second driver, which has a different supporting status to cloud ERP deployment, is cost. The theoretical model suggests that deploying a cloud-based ERP system is cost-efficient; however, in the revised model, the cost factor is no longer positively related to cloud ERP deployment. Case company A has conducted a cost comparison between implementing a cloud ERP and keeping on-premise ERP; the outcome indicates that surprisingly, the two approaches cost equivalently. There is no considerable cost reduction for the case company A. On the other hand, cost-efficient is the major driver for case company B to deploy the cloud ERP system. Hence, a mixed opinion has been drawn on cost driver so that P7 is not supported.

The third driver is vendor reliability. As mentioned earlier, vendor reliability is somehow related to security risks so that in the theoretical model, it is negatively related to cloud ERP deployment. According to the interviewee, as long as there are legal contracts and service-level agreements, vendor reliability is not a serious concern. Therefore, P10 is not supported because a neutral opinion has been provided on vendor reliability.

The other 2 propositions are P3 and P8 where "Customization and Business Complexity are positively related to cloud ERP deployment decisions" are not supported. Because neither perspective (client nor consultant) have shown evidences on the existence of these two drivers.

In addition to that, it is evident that the technological context is the dominating category in the TOE model; this is corresponding with the initial theoretical model where technology is also dominating among the three contexts.

In short, the theoretical model is in agreement with the revised model to an extent; there are 8 out of 10 drivers supported by the revised model, and 5 propositions are supported. Additionally, there are 7 new findings outlined in the revised model; hence, the next section will discuss the emergence of these new findings.

### 4.3.2 Emergence of New Drivers

The revised model (Figure 5) shows that there are seven new drivers identified by both client companies and the consultant company. In order to directly address these new drivers' supporting statuses (supported represents using "Y", not supported represents as "N) from each perspective, Table 7 below is created.

**Table 7**: New Driver Supporting Status by Two Perspectives

| **New Drivers** | **Supporting Status (Y/N)** | |
|---|---|---|
| | **Client** | **Consultant** |
| ND 1: Maturity of ERP systems in the cloud environment | Y | N |
| ND 2: Change of operating model | Y | N |
| ND 3: Vendor Relationship management | Y | N |
| ND 4: Political reason | Y | N |
| ND 5: Functionality | N | Y |
| ND 6: Global expansion | N | Y |
| ND 7: Issues with current system | N | Y |

Based on Table 7, the 4 new drivers from client companies are: maturity of ERP systems in the cloud environment, change of operating model, vendor relationship management, and political reason. Moreover, 3 new drivers from the consultant company are: functionality, issues with the current system, and global expansion.

In general, 5 drivers out of 7 have never been mentioned in any previous scholarly literature, and they are: maturity of ERP systems in the cloud environment, vendor relationship management, political reason, global expansion, and issues with current system. This emergence of these new drivers is profoundly affected by organizational context. Take case company B as an example, vendor relationship management and political reason are two critical drivers for the organization to deploy cloud ERP systems since the company is a state-level enterprise; which means that these two drivers might apply to public organizations. The other three drivers should be well considered, especially for large-scale organizations. Deploying a cloud ERP system without taking any risks, the enterprise will need to ensure the system's maturity and its customer base. Furthermore, as business grows, current system might no longer be suitable for the enterprise, and organizational future expansion is also a critical driver mentioned by the consultant company for both Jak Max case and REA Group case.

As for change of operating model and functionality, these two drivers have already been mentioned in previous studies. Gupta et al. (2017) indicated that organizational change is always hard for enterprises when adopting new systems, especially when employees are already familiar and comfortable with their existing system. Change of operating model, which happens in the transferring stage from on-premise to the cloud, is a part of organizational change; hence, organizational change is an antecedent to change of operating model.

As mentioned earlier, functionality is quite controversial with cloud ERP systems. The consultant company suggests that functionality is an essential driver for enterprises to deploy a cloud ERP system; this is not in agreement with Garverick's study. Garverick (2014) indicated that one of the significant barriers for enterprises to deploy cloud ERP systems is the functionality limitations. However, Johansson et al. (2014) mentioned that cloud-based ERP systems could offer enterprises similar functions just as on-premise ERP provides (p. 2). Duan, Faker, Fesak, and Stuart (2013) have made similar statements; they also believe that cloud ERP offers sufficient functionalities.

In general, these seven new drivers are all supported by different case organizations; each of these drivers should be taken into consideration when making cloud ERP deployment decisions since they might apply in other organizations.

The next section demonstrates the comparison in terms of cloud ERP deployment drivers between client company perspective and the consultant company perspective.

### 4.3.3 Client Company vs Consultant Company Perspective

By looking at those 15 drivers from two perspectives (Figure 5 and Figure 7), it is interesting to see that all of the drivers are supported by either client companies or consultant companies, while none of them are supported by both perspectives.

In the technological context, client companies have mentioned three drivers (e.g. maturity of ERP in the cloud environment, security, and integration) and consultant company mentioned five drivers. Client companies believe that technological related drivers are mainly negatively related to cloud ERP adoption; however, consultant company indicates that by adopting cloud ERP systems, technology efficiency can be guaranteed in terms of the system's scalability, maintainability, accessibility, and functionality.

In the organizational context, four drivers were mentioned by client companies (e.g, cost, business agility, change of operating model, and vendor relationship management); yet, no drivers have been identified from the consultant company perspective. Organizational related drivers are usually the main reasons for enterprises to deploy cloud ERP systems. It is quite intriguing that the consultant company does not indicate cost-efficient as a driver.

In the environmental context, two drivers were mentioned by the client companies (e.g. vendor reliability and political reason) while only one driver was identified by the consultant company (global expansion).

In general, client companies have indicated three drivers which are negatively related to cloud ERP deployment, three drivers positively related, and two drivers have mixed relationships. On the other hand, all drivers identified by the consultant company are positively related to cloud ERP deployment. It is notable that there is not even one common driver identified by both perspectives.

The next section will be concluding the whole chapter.

## 4.4 Conclusion

In conclusion, this whole chapter presents the findings of this research. It includes the interview findings and the outcomes of the secondary resource analysis. In general, 15 drivers have been identified from client companies and the consultant company perspectives; also, the revised model is developed using the 15 drivers.

In this chapter, the revised model has followed the TOE framework and has included two perspectives. Besides, cross-case comparison, comparison between the theoretical model and the revised model, the emergence of seven new drivers, drivers' comparison between two perspectives have all been illustrated.

The next section presents the conclusion of this research including addressing the research questions and limitations of this research.

# Chapter 5: Conclusion

This chapter concludes this research paper and has been divided into four sections. Section 5.1 summarizes the entire thesis. Section 5.2 addresses and answers the research questions. Section 5.3 presents the contributions of this research. The last section 5.4 illustrates the limitation of this research and how this research can be improved in future researches.

## 5.1 Summary

This research intends to understand cloud ERP deployment intentions from both the client company and the consultant company perspectives. In order to address the two research questions, a rigorous systematic literature analysis on cloud ERP has been conducted. The initial theoretical model on "Drivers affecting cloud ERP deployment decisions" which contains 10 drivers has been developed by analysing relevant literature studies.

By adopting a Case Study approach, a semi-structured interview has been conducted to collect primary data from client company perspective in terms of cloud ERP deployment decisions; multiple secondary resources have been used to gather data from the consultant company perspective due to time and access limitations. After analysing those data and compared to the theoretical model, a revised model which includes 15 drivers from both perspectives has been developed. Within those 15 drivers, 8 drivers are retained from the theoretical model and 7 of them are new.

## 5.2 Research Questions

### 5.2.1 Research Question 1

What key drivers influence the deployment decision of cloud ERP systems for the Australian context?

Based on the revised model, 15 drivers were identified from Australian organizations. There are 8 drivers corresponding with the theoretical model and they are security, scalability, accessibility, integration, maintainability, cost, business agility, vendor reliability. Moreover, 7 new drivers have been identified and they are change of operating model, maturity of ERP systems in the cloud environment, vendor relationship management, political reason, functionality, global expansion, and issues with current system. Additionally, within those 7 drivers, "change of operating model" is a moderating driver of "organizational change"; maturing of ERP systems in the cloud environment, vendor relationship management, political reason, issues with current system, and global expansion are new factors which have never been mentioned in any previous scholarly literature.

### 5.2.2 Research Question 2

Do the viewpoints of the client company and the consultant company concerning the influence of key drivers on cloud ERP deployment decision differ?

The viewpoints of client company and consultant company are differentiated from each other in terms of deploying cloud ERP systems. Client companies have plenty of concerns on deploying cloud ERP, many of the drivers they have identified are negatively related to cloud ERP deployment decisions; while for consultant company, all of the drivers are positively related to cloud ERP deployment.

Moreover, not even one common driver has been identified from both perspectives. All the drivers identified by client companies are distinguished from the drivers identified by the consultant company.

## 5.3 Research Contributions

This research makes contributions in both theoretical way and practical way.

In theory, the drivers identified from the revised model suggested that technology is the dominating category from the Technology-Organization-Environment (TOE) framework on cloud ERP deployment decision making in Australia. From research findings, technology is the dominating category; in the theoretical model, technology is also the dominating category. Hence, this contribution might apply to different countries where technology is always the dominating category in terms of cloud ERP deployment. Moreover, the revised model suggested that one moderating driver has been identified which is change of operating model. This moderating driver represents a contribution of knowledge to better understand “cloud ERP deployment” phenomenon.

In practice, a divergence in perception between client companies and consultant companies has been drawn in terms of cloud ERP deployment. This kind of perception represents a significant contribution to IT practice since it will help the formulation of the appropriate customer and consultant relationship. Also, the identified 15 drivers can now be applied in different organizational context to support senior-level managers decision-making on whether to adopt cloud ERP systems.

## 5.4 Limitations and Future Research Directions

This research can be seen as a stepping-stone for future relevant studies. There are several limitations on this research presents as follow.

Firstly, findings are not generalizable for this research due to three reasons. The first reason is that this research has adopted a case study approach as collecting both primary and secondary data. The secondary reason is that this research is conducted only in the Australian context, this finding might not apply to other countries. And the third reason is that only one interview has been conducted due to time and access limitations. Future studies should consider taking multiple case studies (involving multiple interviews or surveys from both perspectives) across different countries so that researches will reach a better understanding of cloud ERP deployment intentions.

Secondly, there are four case organizations involved in this research; yet, all of them are large-scale organizations. According to Johansson et al. (2014), large organizations usually have different viewpoints from small and medium-sized enterprises on cloud ERP deployment decisions. Therefore, the findings of this study only apply to large-size enterprises in the Australian context. Additionally, much research has been conducted in SME cloud ERP adoption. For future work, multiple organizations from different industries in different sizes should be involved so that the findings might be more typical and can be applied in wider ranges.

The third limitation is interview timing. For this research, only one organization (Case company B) has not made the cloud ERP adoption decisions yet; the other three organizations have already made decisions and deployed cloud ERP systems. In this case, drivers within the final model (Drivers influencing cloud ERP deployment model) might differ from Post-deployment and Pre-deployment. Therefore, the findings are limited in terms of interviewing time. Future study should take longitude case study to determine whether the influence of drivers will change. Both pre-deployment and post-deployment investigations can be done to see if there are new drivers emerge. This will provide much richer understandings on cloud ERP deployment phenomena.

The fourth limitation of this research is that the theoretical model was developed based on the previous scholarly literature; hence, the drivers on cloud ERP deployment decisions identified in the theoretical model are not up-to-dated. In addition, the earliest research was conducted in 2010 by Arinze and Anandarajan (Arinze & Anandarajan, 2010); the limitation here is that much of the negativity around cloud ERP systems might has already been resolved to-date such as security issues. Hence, future studies can be established only on recent studies in terms of understanding cloud ERP deployment decisions.

Lat but not least, the research outcome has suggested that a divergence in perceptions of two perspectives on cloud ERP deployment decisions has been drawn; yet, the reason of why these drivers are differentiated from two perspectives need to be analysed in future studies.

**(The End)**

# Appendix A: 31 Literature Category

| **Category** | **Literature** | **Author** |
|---|---|---|
| Benefits / Drawbacks | Cloud ERP: a new dilemma to modern organisations? | Peng, G; Gala, C |
| | ERP in the Cloud–Benefits and Challenges | Lenart, A |
| | Motives and Barriers to Cloud ERP Selection for SMEs: A Survey of Value Added Resellers (VAR) Perspectives | Garverick, M |
| | Potential concerns and common benefits of cloud-based enterprise resource planning (ERP) | Parthasarathy, S |
| | An analysis of the perceived benefits and drawbacks of cloud ERP systems: A South African study | Scholtz, B; Atukwase, D |
| | Benefits and challenges of cloud ERP systems–A systematic literature | Elmonem, M; Nasr, E; Geith, M |
| Migration Journey | Moving from evaluation to trial: How do SMEs start adopting Cloud ERP? | Salim, S; Sedera, D; Sawang, S; Alarifi, A; Hamad, E; Atapattu, M |
| | Switching toward Cloud ERP: A research model to explain intentions | Mezghani, K |
| | Understanding Intentions to Switch Toward Cloud Computing at Firms' Level: A Multiple Case Study in Tunisia | Hachicha, Z; Mezghani, K |
| Lifecycle | Cloud ERP Adoption-A Process View Approach | Salim, S |
| Implementation / Adoption | Identification of challenges and their ranking in the implementation of cloud ERP: A comparative study for SMEs and large organizations | Gupta, S; Misra, S; Singh, A; Kumar, V; Kumar, U |
| | In-house versus in-cloud ERP systems: a comparative study | Elragal, A; El Kommos, M |
| | Factors affecting cloud ERP adoption in Saudi Arabia: An empirical study | AlBar, A; Mustafa H, Md Rakibul |
| | Organizational, technological and extrinsic factors in the implementation of cloud ERP in SMEs | Gupta, S; Misra, S; Kock, N; Roubaud, D |
| | Factors for adopting ERP as SaaS amongst SMEs: The customers vs. Vendor point of view | Rodrigues, J; Ruivo, P; Johansson, B; Oliveira, T |
| | Factors that determine the adoption of cloud computing: A global perspective | Arinze, B; Anandarajan, M |
| | ERP on Cloud: Implementation strategies and challenges | Appandairajan, P; Khan, N; Madiajagan, M |
| | Critical success factors model for business intelligent over ERP cloud | Emam, Ahmed Z |
| | Using the Multi-Theory Approach to Investigate the Factors that Affect the Adoption of Cloud Enterprise Resource Planning Systems by Micro, Small and Medium Enterprises in the Philippines | Caguiat, M; Rowena, M; Suarez, M; Teodosia, C |
| | Cloud ERP Adoption Opportunities and Concerns: A Comparison between SMEs and Large Companies | Johansson, B; Alajbegovic, A; Alexopoulos, V; Desalermos, A |
| | Cloud ERP implementation | Carutasu, N; Carutasu, G |
| | Compliance, network, security and the people related factors in cloud ERP implementation | Gupta, S; Misra, S |
| | Indian SMEs Perspective for election of ERP in Cloud | Mahara, T |
| | Cloud-based ERP solution for modern education in Vietnam | Nguyen, T; Nguyen, T; Misra, S |
| | Competition and challenge on adopting cloud ERP | Weng, F; Hung, M |
| Other | Role of cloud ERP on the performance of an organization: contingent resource-based view perspective | Gupta, S; Kumar, S; Singh, S; Foropon, C; Chandra, C |
| | Cloud ERP simulation in powersim environment | Romanov, V; Varfolomeeva, A |
| | Cloud and traditional ERP systems in small and medium enterprises | Saini, I; Khanna, A; Peddoju, SK |
| | A framework for evaluating cloud enterprise resource planning (ERP) systems | Chandrakumar, T; Parthasarathy, S |
| | Academic Cloud ERP Quality Assessment Model | Surendro, K; Olivia, O |
| | Cloud ERP Meets Manufacturing | Symonds, M |

# Appendix B: 21 Selected Literature

| **Title** | **Author & Date** |
|---|---|
| Academic Cloud ERP Quality Assessment Model | Kridanto Surendro, Olivia, 2016 |
| A Framework for Evaluating Cloud Enterprise Resource Planning (ERP) Systems | T. Chandrakumar and S. Parthasarathy, 2014 |
| An Analysis of the Perceived Benefits and Drawbacks of Cloud ERP Systems: A South African Study | Brenda Scholtz and Denis Atukwase, 2016 |
| Benefits and challenges of cloud ERP systems–A systematic literature review | Elmonem, Mohamed A Abd Nasr, Eman S Geith, Mervat H., 2016 |
| Cloud-Based ERP Solution for Modern Education in Vietnam | Thanh D. Nguyen, Thanh T. T. Nguyen, and Sanjay Misra, 2014 |
| Motives and Barriers to Cloud ERP Selection for SMEs: A Survey of Value Added Resellers (VAR) Perspectives | Michael L. Garverick, 2014 |
| Factors for Adopting ERP as SaaS amongst SMEs: The Customers vs. Vendor Point of View | Rodrigues, Jorge Ruivo, Pedro Johansson, Björn Oliveira, Tiago., 2016 |
| Cloud ERP Adoption Opportunities and Concerns: A Comparison between SMEs and Large Companies | Björn Johansson, Amar Alajbegovic, Vasileios Alexopoulos, Achilles Desalermos, 2014 |
| CLOUD ERP: A NEW DILEMMA TO MODERN ORGANISATIONS? | Peng, Guo Chao Alex Gala, Chirag., 2014 |
| Competition and Challenge on Adopting Cloud ERP | Fumei Weng and Ming-Chien Hung, 2014 |
| Compliance, network, security and the people related factors in cloud ERP implementation | Shivam Gupta and Subhas C Misra, 2016 |
| ERP in the Cloud – Benefits and Challenges | Anna Lenart, 2011 |
| ERP on Cloud: Implementation Strategies and Challenges | P. Appandairajan, Zafar Khan, Madiajagan, 2012 |
| Factors that determine the Adoption of cloud computing: A Global Perspective | Bay Arinze, Murugan Anandarajan, 2010 |
| Factors affecting cloud ERP adoption in Saudi Arabia: An empirical study | AlBar, Adnan Mustafa Hoque, Md Rakibul, 2017 |
| Identification of challenges and their ranking in the implementation of cloud ERP A comparative study for SMEs and large organizations | Shivam Gupta, Subhas Misra, Akash Singh, 2017 |
| Indian SMEs Perspective for election of ERP in Cloud | Tripti Negi Mahara, 2013 |
| In-house versus in-cloud ERP systems: a comparative study | Elragal, Ahmed El Kommos, Malak., 2012 |
| Organisational, technological and extrinsic factors in the implementation of cloud ERP in SMEs | Gupta, Shivam Misra, Subhas C Kock, Ned Roubaud, David., 2018 |
| Potential Concerns and Common Benefi ts of Cloud-Based Enterprise Resource Planning (ERP) | S. Parthasarathy, 2013 |
| Switching Toward Cloud ERP: A Research Model to Explain Intentions | Mezghani, K., 2014 |

# Appendix C: Coding Protocol

**Section A: Paper title and authors**

| | |
|---|---|
| **Title** | |
| **Author** | |
| **Year** | |

**Section B: Research characteristics**

| No. | Brief description of research attributes |
|---|---|
| A1 | Type of research approach<br>1: empirical, 2: conceptual, 3: literature review, 4: methodological |
| A2 | Type of empirical research approach<br>1: experiment, 2: case study, 3: survey, 4: secondary data, 5: design science, 6: mixed, 7: others: _____ |
| A3 | Time dimension of research<br>1: not mentioned, 2: cross-sectional, 3: longitudinal |
| A4 | Mode of data collection<br>1: not mentioned, 2: interview (in-person, online), 3: survey (in-person, postal, online), 6: others: ______ |
| A5 | Participants from who data were collected<br>1: not mentioned, 2: business managers, 3: IT managers, 4: others: CEO |
| A6 | Number of participants from who data were collected<br>1: not mentioned, 2: below 10, 3: between 10 to 49, 4: between 50 to 100, 5: above 100 |
| A7 | Research model<br>1: not mentioned, 2: poorly described, 3: well described |
| A8 | Relevant theory used in research model<br>1: not mentioned, 2: DOI, 3: TOE, 4: TAM, 5: TRA, 6: TPB, 7: UTAUT, 8: RBVF, 9: dynamic capability, 10: transaction cost economics, 11: mixed, 12: others: ______ |
| A9 | Unit of analysis<br>1: not mentioned, 2: individual, 3: organisation, 4: application/system,<br>5: others: ______ |
| A10 | Type of industry<br>1: retail, 2: education, 3: manufacturing, 4: energy, 5: technology,<br>6: others: ______ |
| A11 | Type of organisation<br>1: not mentioned, 2: private, 3: government, 4: both private and government |
| A12 | Size of organisation<br>1: not mentioned, 2: small, 3: medium, 4: large, 5: mixed |
| A13 | Country where the research was done<br>1: not mentioned, 2: Australia, 3: China, 4: USA, 5: UK, 6: India, 7: Africa,<br>8: Japan, 9: others: ______ |
| A14 | From which perspective the research was done<br>1: not mentioned, 2: Vendor, 3: Consultant company, 4: Client company,<br>5: none |

**Section B: Key areas of cloud ERP adoption**

| No. | Brief description of research attributes |
|---|---|
| B1 | Definition of cloud ERP<br>1: not mentioned, 2: own definition, 3: referred to a definition provided by other scholars |
| B2 | Deploymet factors mentioned<br>1: No, 2: Yes |
| B3 | List of supported factors<br>1: Security, 2: Scalability, 3: Customization, 4: Maintainability, 5: Integration,<br>6: Network; 7: Cost, 8: Business complexity, 9: Organisational Change,<br>10: Vendor reliability, 11: top management support, 12: compatibility,<br>13: other: ______ |
| B4 | Type of influence on factors<br>1: not mentioned, 2: negative, 3: positive, 4: mixed |
| B5 | Number of factors considered in the paper<br>1: not mentioned, 2: less than 5, 3: between 5 to 10, 4: 10 and above |

# Appendix D: Drivers Affecting Cloud ERP Deployment

| Category | Drivers |
|---|---|
| Technological (N = 43) | Functionality (Parthasarathy, 2013)<br>Limited functionality (Garverick, 2014)<br>Performance (Scholtz et al., 2016)<br>Deployment speed (Johansson et al., 2014)<br>Security (Arinze & Anandarajan, 2010; Lenart, 2011; Appandairajan et al., 2012; Elragal et al., 2012; Chandrakumar & Parthasarathy, 2014; Garverick, 2014; Johansson et al., 2014; Mezghani, 2014; Nguyen et al., 2014; Peng & Gala, 2014; Weng & Hung, 2014; Scholtz et al., 2016; Surendro et al., 2016; Rodrigues et al., 2016)<br>Stability (Arinze & Anandarajan, 2010;)<br>Rapid deployment (Lenart, 2011)<br>Data ownership (Lenart, 2011)<br>Application ownership (Lenart, 2011)<br>Lower capacity requirements (Lenart, 2011)<br>Scalability (Lenart, 2011; Appandairajan et al., 2012; Elragal et al., 2012; Mahara, 2013; Parthasarathy, 2013; Peng & Gala, 2014; Surendro et al., 2016; Gupta et al., 2017)<br>Flexibility (Appandairajan et al., 2012; Parthasarathy, 2013; AlBar & Hoque, 2014)<br>Accessibility (Appandairajan et al., 2012; Elragal et al., 2012; Mahara, 2013; Mezghani, 2014)<br>Availability (Chandrakumar & Parthasarathy, 2014; Surendro et al., 2016; Rodrigues et al., 2016; Mahara, 2013)<br>Integration (Elragal et al., 2012; Chandrakumar & Parthasarathy, 2014; Garverick, 2014; Mezghani, 2014; Scholtz et al., 2016; Rodrigues et al., 2016)<br>Data backup (Surendro et al., 2016)<br>Easy upgrades (Lenart, 2011)<br>Interoperability (Mahara, 2013)<br>Maintainability (Parthasarathy, 2013; Chandrakumar & Parthasarathy, 2014)<br>Customization (Elragal et al., 2012; Garverick, 2014)<br>Mobility (Mahara, 2013; Parthasarathy, 2013)<br>Usability (Chandrakumar & Parthasarathy, 2014)<br>Compatibility (Chandrakumar & Parthasarathy, 2014)<br>Rapid response (Elmonem et al., 2016)<br>Compliance (Nguyen et al., 2014)<br>Network (Nguyen et al., 2014) |
| | Downtime (Weng & Hung, 2014)<br>Higher utilization (Elmonem et al., 2016)<br>Time-saving (AlBar & Hoque, 2017)<br>Data integrity (Rodrigues et al., 2016)<br>Innovation (Nguyen et al., 2014)<br>Better IT support (Peng & Gala, 2014)<br>Observability (AlBar & Hoque, 2017)<br>Upgrade & Enhancement (Peng & Gala, 2014)<br>ICT Skills (AlBar & Hoque, 2017)<br>ICT infrastructure (AlBar & Hoque, 2017)<br>Risk of open access (Lenart, 2011)<br>Data backup and recovery (Mahara, 2013)<br>Lack of control (Mahara, 2013)<br>IT Resources (Johansson et al., 2014)<br>Ubiquity (Rodrigues et al., 2016)<br>Data extraction (Gupta et al., 2017)<br>Resource isolation (Surendro, 2016) |
| Organizational (N = 30) | Business Support (Arinze & Anandarajan, 2010)<br>Cost (Lenart, 2011; Appandairajan et al., 2012; Elragal et al., 2012; Mahara, 2013; Parthasarathy, 2013; Chandrakumar & Parthasarathy, 2014; Garverick, 2014; Johansson et al., 2014; Mezghani, 2014; Peng & Gala, 2014; Weng & Hung, 2014; Elmonem et al., 2016; Surendro et al., 2016; AlBar & Hoque, 2017; Rodrigues et al., 2016; Gupta et al., 2017)<br>Business efficiency (Appandairajan et al., 2012)<br>Flexible payment (Mahara, 2013)<br>Organizational culture (AlBar & Hoque, 2017)<br>Loss of IT competencies (Appandairajan et al., 2012)<br>Less staff (Mahara, 2013)<br>Lower upfront cost (Elmonem et al., 2016)<br>Lower operating cost (Elmonem et al., 2016)<br>Lower software cost (Nguyen et al., 2014)<br>Organizational change (Peng & Gala, 2014)<br>Business process (Lenart, 2011)<br>Attitude (Mezghani, 2014)<br>Project champion (Parthasarathy, 2013)<br>Subjective norms (Mezghani, 2014)<br>Organizational culture (AlBar & Hoque, 2017)<br>Behaviour control (Mezghani, 2014)<br>Business process re-engineering (Misra et al., 2018)<br>Business continuity (Johansson et al., 2014)<br>Concentrate on core business (Parthasarathy, 2013) |
| | Business complexity (Garverick, 2014; Scholtz et al., 2016)<br>User involvement (Nguyen et al., 2014)<br>Project team (Nguyen et al., 2014)<br>Top management support (Nguyen et al., 2014)<br>User training (Nguyen et al., 2014)<br>Manageability (Peng & Gala, 2014)<br>Business Agility (Arinze & Anandarajan, 2010; Gupta et al., 2017)<br>Core competencies (Elmonem et al., 2016)<br>Communication (Misra et al., 2018) |
| Environmental (N = 6) | Vendor reliability (Parthasarathy, 2013; Garverick, 2014; Elmonem et al., 2016)<br>Vendor lock-in (Peng & Gala, 2014)<br>Regulatory environment (AlBar & Hoque, 2017)<br>Trust in vendors (Chandrakumar & Parthasarathy, 2014; Nguyen et al., 2014; Rodrigues et al., 2016)<br>Vendor Selection (Johansson et al., 2014; Nguyen et al., 2014)<br>Vendor integrity (Peng & Gala, 2014; Scholtz et al., 2016) |
| **Total** | **79** |